\documentclass[prd, 10pt,aps,nofootinbib, floatfix, notitlepage,twocolumn,longbibliography, superscriptaddress]{revtex4-1}

\usepackage{amsmath,amsfonts,amssymb,mathrsfs} 
\usepackage{graphicx}
\usepackage{dcolumn}
\usepackage{array}
\usepackage{bm}
\usepackage{xcolor}

\PassOptionsToPackage{colorlinks=true,citecolor=blue,urlcolor=blue}{hyperref}
\usepackage{orcidlink}
\usepackage[english]{babel}
\usepackage{url}
\usepackage{adjustbox}
\usepackage[utf8]{inputenc}

\def\bs{\boldsymbol}

\newcommand{\cph}[1]{\textcolor{red}{[cite]}}
\begin{document}

\title{Sensitivity of Weak Lensing Surveys to Gravitational Waves from Inspiraling Supermassive Black Hole Binaries}

\author{Tal Adi\,\orcidlink{0000-0002-5763-9353}}
\email{taladi@usc.edu}
\affiliation{University of Southern California, Los Angeles, CA 90089, USA}

\author{Kris Pardo\,\orcidlink{0000-0002-9910-6782}}
\affiliation{University of Southern California, Los Angeles, CA 90089, USA}

\author{Olivier Doré\,\orcidlink{0000-0001-7432-2932}}
\affiliation{Jet Propulsion Laboratory, California Institute of Technology, 4800 Oak Grove Drive, Pasadena, CA 91109, USA}
\affiliation{California Institute of Technology, 1200 E. California Boulevard, Pasadena, CA 91125, USA}

\begin{abstract}

We explore the sensitivity of weak lensing surveys to gravitational waves (GWs) emitted by inspiraling supermassive black hole binaries (SMBHBs) in the nanohertz to microhertz frequency band, bridging the gap between pulsar timing arrays and space-based interferometers. Building on the formalism for GW-induced shear distortions, we develop a signal-to-noise framework that incorporates survey characteristics such as cadence, angular resolution, and depth. We model the effective galaxy population to evaluate the noise power spectral density and derive characteristic strain sensitivity curves.
Applying this framework to both LSST-like and idealized survey configurations, we show that current surveys are limited by angular resolution and measurement noise, while an idealized, cosmic-variance-limited survey could in principle probe this frequency range. We emphasize that such sensitivity requires observational capabilities far beyond those of existing or planned facilities, and our results should be interpreted as an ultimate limit on the information accessible through weak lensing measurements.
\end{abstract}

\maketitle
 
%%%%%%%%%%%%%%%%%%%%%%%%%%%%%%%%%%%%%%%%%%%%%%%%%%%%%%%%%%%
\section{Introduction}
%%%%%%%%%%%%%%%%%%%%%%%%%%%%%%%%%%%%%%%%%%%%%%%%%%%%%%%%%%%

Gravitational wave (GW) astronomy has opened a new window into the universe, allowing us to probe astrophysical phenomena and fundamental physics in unprecedented ways~\cite{LIGOScientific:2016aoc,Maggiore:1999vm}. The increasing sensitivity of ground-based detectors such as LIGO, Virgo, and KAGRA~\cite{LIGOScientific:2014pky,VIRGO:2014yos,KAGRA:2020tym} has led to the detection of numerous compact binary mergers~\cite{LIGOScientific:2025slb}, providing insights into the population properties of black holes~\cite{LIGOScientific:2025pvj}, and recently even the testing of general relativity in the strong-field regime~\cite{LIGOScientific:2025rid,LIGOScientific:2025brd}.

While ground-based detectors are sensitive to GWs in the frequency range of tens to thousands of Hertz, other methods are required to probe lower-frequency GWs. The upcoming LISA~\cite{LISA:2017pwj}, TianQin~\cite{TianQin:2015yph}, and DeciGO~\cite{Kawamura:2018esd} space missions will target the millihertz and decihertz regimes, where massive black hole binaries and extreme mass ratio inspirals are expected to be prominent sources~\cite{Baker:2019nia}. At even lower frequencies, pulsar timing arrays (PTAs)~\cite{EPTA:2023fyk,NANOGrav:2023gor,Zic:2023gta} are sensitive to nanohertz GWs, primarily from supermassive black hole binaries (SMBHBs)~\cite{NANOGrav:2023hfp}, and stochastic GW backgrounds (SGWBs)~\cite{Sesana:2008mz,Rosado:2015epa,Siemens:2019xkk}. However, there remains a frequency gap between the sensitivity ranges of PTAs and space-based detectors like LISA, roughly spanning $10^{-7}$ to $10^{-4}$ Hz, where only a limited number of detection methods have been proposed~\cite{Armstrong:2003ay,Fedderke:2021kuy,Bustamante-Rosell:2021daj,Sesana:2019vho}.

A primary motivation for exploring this frequency range is the population of SMBHBs. Formed through galaxy mergers and expected to reside in galactic centers~\cite{Begelman:1980vb}, these systems emit GWs throughout their long inspiral. In the nanohertz band probed by PTAs, the most massive and widely separated binaries evolve extremely slowly, producing signals that are effectively monochromatic over observational timescales. At higher frequencies, approaching the microhertz regime, there is more rapid orbital evolution, and the largest systems can no longer be treated as strictly monochromatic, instead exhibiting measurable frequency drift. Individually resolvable SMBHBs may therefore appear as continuous or slowly chirping sources across the nanohertz–microhertz band, while the ensemble of unresolved systems produces a stochastic GW background (SGWB) whose amplitude and spectral shape depend on their merger rate, masses, and dynamical environments~\cite{Phinney:2001di,Burke-Spolaor:2018bvk}.
Recent PTA observations reporting evidence for a nanohertz SGWB~\cite{NANOGrav:2023gor,EPTA:2023fyk} further motivate extending gravitational-wave sensitivity beyond the PTA band. In particular, several beyond-standard-model explanations proposed for the PTA signal predict features that peak or turn over at microhertz frequencies~\cite{NANOGrav:2023hvm}, while at the low-frequency end of the microhertz band the signal is expected to be dominated by the local SMBHB population, dominated by binaries at moderate redshift ($z \sim 1\text{–}2$), offering a direct probe of the nearby number density and mass distribution of these systems~\cite{Sesana:2008mz}.

Given the importance of accessing this microhertz frequency range, it is valuable to explore additional methods capable of detecting gravitational waves in the PTA–LISA gap. Astrometric approaches aim to measure the correlated deflections of apparent stellar positions induced by passing gravitational waves in wide-field photometric surveys~\cite{Book:2010pf,Darling:2018hmc}. More recently, relative astrometry has been proposed as a practical realization of this idea, in which GW-induced correlations are extracted from differential position measurements, allowing general photometric surveys to be used for the measurement~\cite{Wang:2022sxn,Pardo:2023cag,Zhang:2025srs}.

More generally, any gravitational wave passing through the Earth perturbs photon trajectories between distant sources and the observer. This leads not only to timing residuals in pulsar time-of-arrival measurements and correlated deflections in apparent stellar positions, but also to distortions in observed redshifts and in the shapes of extended astronomical objects, such as galaxies~\cite{Schmidt:2012nw,Schmidt:2012ne,Jeong:2012nu,Schmidt:2013gwa}. More recently, these ideas have been extended to explicitly time-dependent observables, with proposals to detect gravitational waves through the temporal evolution of cosmic shear in wide-field imaging surveys~\cite{Mentasti:2024fgt}.

Next-generation cosmic-shear programs are rapidly advancing, with surveys such as the \emph{Vera C. Rubin Observatory's Legacy Survey of Space and Time} (LSST)~\cite{LSSTDarkEnergyScience:2022nlw}, the \emph{Nancy Grace Roman Space Telescope}~\cite{Spergel:2015sza}, and \emph{Euclid}~\cite{EUCLID:2011zbd} set to map billions of galaxies over large fractions of the sky with high angular resolution and depth. These surveys will provide exquisite measurements of weak gravitational lensing signals, enabling precise cosmological constraints~\cite{Euclid:2024yrr,Zeghal:2024kic,Bera:2025ixc}. The same datasets can also be exploited--albeit with considerable observational challenges--to search for GW-induced shear distortions, potentially opening a new avenue for GW detection in the nanohertz to microhertz frequency range.

In this work, we build upon the formalism introduced in Refs.~\cite{Schmidt:2012nw,Mentasti:2024fgt} for the SGWB, to estimate the sensitivity of weak lensing surveys to GWs from inspiraling SMBHBs. We develop a signal-to-noise ratio (SNR) framework that incorporates realistic survey parameters, such as depth, angular resolution, and cadence, and extends the noise treatment in Ref.~\cite{Mentasti:2024fgt} by explicitly relating galaxy number density and size to survey depth. Applying this framework, we assess the sensitivity of forthcoming surveys to GWs from SMBHBs and explore an idealized configuration that probes the performance limits of the method, providing complementary coverage across the nanohertz-to-microhertz band. We also apply this realistic framework to the SGWB sensitivity developed in previous work.

This paper is organized as follows. In Sec.~\ref{sec:methodology}, we review the GW-induced shear distortion formalism and develop the SNR framework for weak lensing surveys. We then review the noise model and present a formalism for estimating the effective number of galaxies as a function of survey depth. In Sec.~\ref{sec:results}, we present sensitivity estimates for LSST-like and idealized survey configurations, and discuss the implications for GW detection. Finally, in Sec.~\ref{sec:conclusion}, we summarize our findings and outline prospects for future work.

%%%%%%%%%%%%%%%%%%%%%%%%%%%%%%%%%%%%%%%%%%%%%%%%%%%%%%%%%%%
\section{Methodology}\label{sec:methodology}
%%%%%%%%%%%%%%%%%%%%%%%%%%%%%%%%%%%%%%%%%%%%%%%%%%%%%%%%%%%

In this section, we present the formalism used to compute the sensitivity of weak lensing surveys to GWs from inspiraling binaries. We begin by reviewing the GW-induced distortions formalism introduced in Ref.~\cite{Mentasti:2024fgt}, which describes how GWs perturb the apparent shapes of extended astronomical objects. We then develop a signal-to-noise ratio (SNR) framework for estimating the sensitivity of a survey with $N$ galaxies to GW signals, incorporating the relevant noise sources and survey parameters. Finally, we present models for the shear measurement error and the number of galaxies observable in a survey as functions of its characteristics.

%%%%%%%%%%%%%%%%%%%%%%%%%%%%%%%%%%%%%%%%%%%%%%%%%%%%%%%%%
\subsection{GW-Induced Shape Distortions}\label{sec:gw_distortion}
%%%%%%%%%%%%%%%%%%%%%%%%%%%%%%%%%%%%%%%%%%%%%%%%%%%%%%%%

We begin by summarizing the formalism introduced by Mentasti \& Contaldi~\cite{Mentasti:2024fgt}, which describes how a passing GW induces small, time-dependent distortions in the apparent shapes of extended astronomical objects. This framework closely parallels that of weak gravitational lensing~\cite{Dodelson:2020bqr}, but here the perturbations are tensorial rather than scalar.

Let $I(\hat n)$ denote the observed flux profile of an image on the sky, where $I$ is the intensity measured along the direction $\hat n$ on the celestial sphere. If the image is distorted relative to its true, undistorted appearance, the observed and true profiles are related by a small angular displacement $\delta\hat n$ of the photon arrival direction:
\begin{equation}
    I_{\mathrm{obs}}(\hat n)
    = I_{\mathrm{true}}(\hat n + \delta\hat n).
    \label{eq:Iobs_Itrue}
\end{equation}
The deflection $\delta\hat n$ encodes the cumulative effect of metric perturbations along the line of sight. For small distortions ($|\delta\hat n|\ll1$), it is convenient to describe the mapping between the true and observed images through a linear operator, the \emph{distortion matrix}
\begin{equation}
    A_{ab}
    \equiv
    \frac{\partial(\hat n_a + \delta\hat n_a)}{\partial \hat n_b}
    \equiv \delta_{ab} + \psi_{ab}.
    \label{eq:Aab}
\end{equation}
Indices $a,b=\{\theta,\phi\}$ span the local tangent plane to the image centroid. The distortion matrix can be decomposed into irreducible components with respect to two-dimensional rotations as
\begin{equation}
     \psi_{ab} \equiv - \kappa\,\delta_{ab} + \epsilon_{ab}\,\omega + S_{ab},
    \label{eq:psi_ab_decomp}
\end{equation}
where $\kappa$ is the \emph{convergence}, $\omega$ the \emph{rotation}, and $S_{ab}$ a symmetric, trace-free tensor describing \emph{shear}. The components of $S_{ab}$ are conventionally expressed as
\begin{equation}
    S_{ab} =
    -\begin{pmatrix}
        \gamma_1 & \gamma_2 \\
        \gamma_2 & -\gamma_1
     \end{pmatrix},
\end{equation}
so that the four quantities $(\kappa, \omega, \gamma_1, \gamma_2)$ completely specify the image distortion. These quantities are mathematically analogous to the Stokes parameters $(I,V,Q,U)$ that describe the polarization of light~\cite{Jackson:1998nia}. It is convenient to combine the two shear components into a single \emph{complex shear parameter} defined as
\begin{equation}
    {}_{\pm2}\gamma = \gamma_1 \pm i\,\gamma_2.
    \label{eq:complex_shear}
\end{equation}
The quantity ${}_{\pm2}\gamma$ transforms as a spin-$\pm2$ field on the celestial sphere: under a right-handed rotation by an angle~$\alpha$ about the line of sight, ${}_{\pm2}\gamma \rightarrow {}_{\pm2}\gamma\, e^{\mp 2i\alpha}$. This spin-weighted formalism parallels that used for the Stokes parameters $Q \pm iU$ in CMB polarization analysis, and allows us to express the GW response functions $F_{{}_{\pm2}\gamma}$ in Eq.~\eqref{eq:FX} in a compact and rotation-covariant form.

A GW induces the deflection of a photon trajectory through the metric perturbation $h_{ij}$. For a wave propagating along the unit direction $\hat{q}\left(\theta,\phi\right)$, the astrometric deflection of a line of sight $\hat{n}\left(\theta_s,\phi_s\right)$ is given by~\cite{Book:2010pf,Mihaylov:2018uqm}
\begin{equation}
    \delta \hat n_i(\hat n, \hat q)
    = \frac{1}{2}
      \!\left[
        \frac{\hat n_i + \hat q_i}{1+\hat q\!\cdot\!\hat n}
        h_{jk}(\hat q)\,\hat n^j\hat n^k
        - h_{ij}(\hat q)\,\hat n^j
      \right].
    \label{eq:deflection}
\end{equation}
This expression corresponds to the ``Earth term'' of the deflection, valid in the limit where the source distance greatly exceeds the GW wavelength\footnote{``Star term" is suppressed by a factor of $1/d$, where $d$ is the distance to the light source, making it irrelevant~\cite{Book:2010pf,Madison:2020xhh}}.
Since the vectors $\hat q$ and $\hat n$ are unit vectors, the angular deflection induced by the gravitational wave is of order the strain amplitude $h$, and is independent of the distance to the source~\cite{Book:2010pf}.
Differentiating Eq.~\eqref{eq:deflection} with respect to $\hat n$ yields the three-dimensional distortion tensor $\tilde{\psi}_{ij} = \partial(\delta\hat n_i)/\partial\hat n_j$.
Projecting this tensor onto the local tangential basis $\{ \hat e_\theta, \hat e_\phi \}$ produces the observable two-dimensional distortion matrix,
\begin{equation}\label{eq:psi_ab}
    \psi_{ab}(\hat q, \hat n)
    = e_a^{~i}(\hat n)\,\tilde{\psi}_{ij}(\hat q,\hat n)\,e_b^{~j}(\hat n),
\end{equation}
whose components correspond to the quantities $(\kappa,\omega,{}_{\pm2}\gamma)$ defined above.

To connect the GW-induced deflection to observable quantities in weak lensing surveys, we decompose the effect into distortions of the apparent shapes of background galaxies. These distortions can be described by dimensionless quantities, the \emph{response functions} $F_\kappa$, $F_\omega$, and $F_{{}_{\pm2}\gamma}$, that encode how the GW direction and polarization map onto observable shear patterns on the sky,
\begin{equation}
    \tilde{\psi}_{ab}\equiv\left(\begin{array}{cc}
-F_{\kappa}-F_{\gamma_{1}} & -F_{\gamma_{2}}+F_{\omega}\\
-F_{\gamma_{2}}-F_{\omega} & -F_{\kappa}+F_{\gamma_{1}}
\end{array}\right).
\end{equation}
It was shown, in Ref.~\cite{Mentasti:2024fgt}, that for a unit-amplitude, left-circularly polarized plane wave, propagating in the $\hat{z}$ direction, these functions take the compact form
\begin{align}
    F_\kappa^L \left(\hat{z},\hat{n}\right) &= -\tfrac{1}{2}\,e^{-2i(\phi - \phi_s)}(1-\cos\theta_s), \nonumber\\
    F_\omega^L \left(\hat{z},\hat{n}\right) &=  -i\,F_\kappa^L\left(\hat{z},\hat{n}\right), \label{eq:FX}\\
    F_{{}_{+2}\gamma}^L \left(\hat{z},\hat{n}\right) &= e^{-2i(\phi - \phi_s)}, \nonumber\\
    F_{{}_{-2}\gamma}^L \left(\hat{z},\hat{n}\right) &=  0, \nonumber
\end{align}
where $(\theta_s,\phi_s)$ are the spherical coordinates of the source direction $\hat n$, and $\phi$ is the azimuthal angle of $\hat{q}$, which is arbitrarily chosen for a wave propagating along $\hat{q}=\hat{z}$. It is straightforward to verify that the response functions for a right-circularly polarized wave are given by $F_X^R = (F_X^L)^*$ for $X \in \{\kappa, \omega, {}_{\pm2}\gamma\}$\footnote{Note that $F_{{}_{\pm2}\gamma}^* = F_{{}_{\mp2}\gamma}$.}.
Eq.~\eqref{eq:FX} can be generalized to an arbitrary GW propagation direction $\hat q$ by performing a suitable rotation of the coordinate system $R(\theta,\phi)$, namely $\hat{q}=R\cdot\hat{z}$ and $\hat{n}'=R\cdot\hat{n}$, under which the response functions transform as spin-weighted fields:
\begin{align}
    F_\kappa^L \left(\hat{z},\hat{n}\right) &= F_\kappa^L \left(\hat{q},\hat{n}'\right), \nonumber\\
    F_\omega^L \left(\hat{z},\hat{n}\right) &=  F_\omega^L\left(\hat{q},\hat{n}'\right), \label{eq:FX_rot}\\
    F_{{}_{\pm2}\gamma}^L \left(\hat{z},\hat{n}\right) &= F_{{}_{\pm2}\gamma}^L\left(\hat{q},\hat{n}'\right)e^{\mp2i\phi}. \nonumber
\end{align}
Eq.~\eqref{eq:FX} encapsulates the angular pattern of the GW-induced distortions and serves as the starting point for the signal-to-noise formalism developed in the following sections. A complete re-derivation is provided in Appendix~\ref{app:gw_dist}.

%%%%%%%%%%%%%%%%%%%%%%%%%%%%%%%%%%%%%%%%%%%%%%%%%%%%%%%%%%
\subsection{Measurement Noise in Time-Varying Shear}\label{sec:noise_model}
%%%%%%%%%%%%%%%%%%%%%%%%%%%%%%%%%%%%%%%%%%%%%%%%%%%%%%%%%%

The shear noise in weak lensing measurements arises from two main sources: the shape noise due to the intrinsic ellipticity distribution of galaxies, and the measurement error associated with estimating the shear from noisy images (e.g., Ref.~\cite{Chang:2013xja}),
\begin{equation}
    \sigma_{\gamma,i}^2 = \sigma_{SN}^2 + \sigma_{m,i}^2. 
\end{equation}
In our case, however, the observable is the \emph{temporal variation} of the measured distortion. Since the intrinsic ellipticity of a galaxy is static in time, differencing observations across multiple epochs removes the shape noise contribution. As a result, the sensitivity is determined primarily by the measurement uncertainty in galaxy shapes, $\sigma_{m,i}$.

To estimate $\sigma_{m,i}$, we follow Ref.~\cite{Mentasti:2024fgt} and model the measurement error as the differences between the resolved and observed distortions,
\begin{equation}
    \sigma_{ij}=\frac{\left|q_{ij}^\mathrm{res} - q_{ij}\right|}{q_{ij}}.
\end{equation}
Here, $q_{ij}$ are the second moments of the galaxy's image brightness distribution, defined as
\begin{equation}
    q_{ij}\equiv\left\langle \theta_{i}\theta_{j}\right\rangle _{I_{\text{obs}}}\equiv\frac{\int d^{2}\theta\,I_{\text{obs}}\left(\bs{\theta}\right)\theta_{i}\theta_{j}}{\int d^{2}\theta\,I_{\text{obs}}\left(\bs{\theta}\right)},
\end{equation}
where $\left\langle \cdot\right\rangle _{I_{\text{obs}}}$ denotes the brightness-weighted average over the observed image $I_{\text{obs}}(\bs{\theta})$, with $\bs{\theta}=(\theta_1,\theta_2)$ being angular coordinates on the sky. The resolved second moments $q_{ij}^\mathrm{res}$ are computed similarly, assuming finite angular resolution $\sigma_\theta$, such that
\begin{equation}
    q_{ij}^{\mathrm{res}}=\frac{\sigma_{\theta}^{2}}{I_{0}\Theta^{2}}\sum_{n=1}^{N_{\text{pix}}}I_{\text{obs}}\left(\boldsymbol{\theta}^{n}\right)\theta_{i}^{n}\theta_{j}^{n},
\end{equation}
where, for simplicity, we also assumed that the intensity $I_{\text{obs}}$ is constant over the covered area of the observed object, $\Theta$. As shown in Ref.~\cite{Mentasti:2024fgt}, the measurement error scales with the angular resolution as
\begin{equation}\label{eq:measurement_error}
    \sigma_{X} \sim \sigma_{ij} \simeq \frac{\sigma_\theta^2}{\Theta^2}.
\end{equation}

%%%%%%%%%%%%%%%%%%%%%%%%%%%%%%%%%%%%%%%%%%%%%%%%%%%%%%%%%%
\subsection{Survey-Averaged SNR and Sensitivity}\label{sec:snr_formalism}
%%%%%%%%%%%%%%%%%%%%%%%%%%%%%%%%%%%%%%%%%%%%%%%%%%%%%%%%%%

To estimate the signal-to-noise ratio (SNR) for a survey with $N$ galaxies, we first consider the observable distortion measurement for the $i$-th galaxy, defined as follows:
\begin{equation}
    s_X^i(t)=X^i(t) + X_0^i +n_X^i(t),
\end{equation}
where
\begin{equation}
    X^i(t)=\sum_{p}F_{X}^{p}\left(\hat{q},\hat{n}^i\right)h_{p}\left(t\right)
\end{equation}
is the response to the GW, with $p\in\left\{ R,L\right\}$ denoting the right and left polarizations, and $n^i_X(t)$ is the detector noise. $X^i_0$ is the constant, time-independent contribution to the distortion, which we assume can be perfectly subtracted from the data.

It is convenient to work in Fourier space, where the noise is characterized by a power spectral density (PSD). For simplicity, we assume that the noise in each galaxy is Gaussian, white, and uncorrelated between galaxies, with a single-sided PSD $S_n^X$ defined by
\begin{equation}\label{eq:noise_psd}
    \left\langle \tilde{n}_X^i(f)\tilde{n}_X^{j*}(f')\right\rangle = \frac{1}{2}\delta_{ij}\delta(f-f')S_n^X,
\end{equation}
where $\tilde{n}_X^i(f)$ is the Fourier transform of $n_X^i(t)$.

The variance of the time-domain noise can then be expressed in terms of the PSD as
\begin{equation}
    \sigma_X^2 = \langle n_X^2(t)\rangle = \int_{1/T}^{1/(2\Delta t)} S_n^X\,df \simeq \frac{S_n^X}{2\Delta t},
\end{equation}
where $T$ is the total observation time and $\Delta t$ is the cadence.

As discussed in the previous section, the relevant noise contribution arises from the measurement uncertainty in galaxy shapes. We therefore identify this variance with the measurement error $\sigma_{X,m}^2$ in Eq.~\eqref{eq:measurement_error}, establishing the connection between the observational uncertainty and the effective noise PSD $S_n^X$.

Under these assumptions, the optimal matched-filter SNR for a single galaxy $i$ and distortion component $X$ is given by
\begin{align}
    \left(\frac{S_X}{N_X}\right)^2  (\hat q,\hat n)&=
        4\int_0^\infty
        \frac{\left|\tilde{X}(f)\right|^2}{S_n^X}\,df \label{eq:SNR2}\\
    &= 4\int_{0}^{\infty}\frac{\left|\sum_{p}F_{X}^{p}\left(\hat{q}, \hat{n}\right)\tilde{h}_{p}\left(f\right)\right|^{2}}{S_{n}^{X}\left(f\right)}\,df, \nonumber
\end{align}
where $\tilde{h}_p(f)$ is the Fourier transform of $h_p(t)$, and we dropped the index $i$ for brevity. 

For a GW wave from a inspiraling binary, as described in Appendix~\ref{app:waveform}, Eq.~\eqref{eq:SNR2} becomes
\begin{equation}
    \left(\frac{S_X}{N_X}\right)^2 (\hat q,\hat n') = \frac{8}{5}\int_{0}^{\infty}
    \frac{|F_X^L(\hat z,\hat n)|^2 \left|\tilde{h}(f)\right|^2}
         {S_n^X}\,df,
    \label{eq:SNR2_final}
\end{equation}
where $\tilde{h}(f)$ is the Fourier transform of the strain amplitude of the form $h(t)=h_0\cos\left[\Phi(t)\right]$.
Here we used the spin symmetry from Eq.~\eqref{eq:FX_rot} to rewrite the SNR in terms of the rotated source direction $\hat n$. We then average over the binary's inclination and polarization angles and exploit the symmetry between the left- and right-circular responses; this removes cross terms between polarizations and yields the overall factor $8/5$ (see Appendix~\ref{app:matched_filter}). Lastly, we average the squared SNR over the full sky, and sum over all distortion components, assuming uniform noise PSD $S_n^X=S_n$ for all components, which gives
\begin{align}
    \left\langle\left(\frac{S}{N}\right)^{2}\right\rangle &=\sum_{X}\left\langle\left(\frac{S_X}{N_X}\right)^{2}\right\rangle \label{eq:SNR2_avg}\\
    &=\frac{28}{15} \int_{0}^{\infty}
    \frac{ \left|\tilde{h}(f)\right|^2}
         {S_n}\,df. \nonumber
\end{align}
Here, $\langle\cdot\rangle$ denotes averaging over the sky, binary orientation, and polarization angles.
We assume statistically independent noise between different galaxies, so that the total SNR for a survey with $N$ galaxies is given by
\begin{equation}
    \left(\frac{S}{N}\right)_{\mathrm{total}}^{2} = \sum_N\frac{28}{15} \int_{0}^{\infty}
    \frac{ \left|\tilde{h}(f)\right|^2} {S_n}\,df.
    \label{eq:SNR2_total}
\end{equation}

As a first approximation, one may consider a typical angular size for all galaxies in the survey, such that the noise PSD $S_n$ is the same for all galaxies. In this case, the total SNR simply scales as $\sqrt{N}$ compared to that of a single galaxy. However, as discussed in Sec.~\ref{sec:noise_model}, the measurement noise PSD, $S_n$, depends on the angular size of each galaxy, which varies across the survey population. A more accurate treatment would therefore involve summing over the individual noise PSDs for each galaxy, $S_n^i$, leading to
\begin{equation}
    \left(\frac{S}{N}\right)_{\mathrm{total}}^{2} = 4\int_{0}^{\infty}\frac{\left|\tilde{h}\left(f\right)\right|^{2}}{S_{n}^{\text{eff}}}df,
    \label{eq:SNR2_total_Sn_eff}
\end{equation}
where we defined an effective noise PSD for the survey,
\begin{equation}\label{eq:Sn_sigma}
    S_n^{\mathrm{eff}} \equiv\frac{15}{7} \left[\sum_{i=1}^N\frac{1}{S_n}\right]^{-1} \simeq \frac{30}{7}\Delta t\left[\sum_{i=1}^N\left(\frac{\Theta^2}{\sigma_\theta^2}\right)^2\right]^{-1}.
\end{equation}

These expressions make clear that the SNR scales linearly with the GW strain amplitude and, like the angular deflection itself, is independent of the distance to the source. The noise—and thus the sensitivity—is therefore primarily determined by the ability to measure small, time-dependent distortions in galaxy shapes, as encapsulated in Eq.~\eqref{eq:measurement_error}. Importantly, while the GW-induced shear signal is independent of the angular size of the source, the measurement noise depends on it through the ability to resolve and accurately characterize galaxy shapes.

We note that this model assumes an idealized scenario, neglecting other potential sources of measurement error, such as photon shot noise, background noise, and systematic effects. A more comprehensive treatment would require detailed simulations of the imaging process and shape measurement algorithms (e.g., \cite{Chang:2013xja}).

It is instructive to define the \emph{effective} number of galaxies per solid angle as
\begin{equation}\label{eq:n_eff_def}
    \tilde{n}_{\text{eff}}\equiv\frac{1}{\Omega}\sum_{i=1}^N\frac{\sigma_{0}^{2}}{\sigma_{m,i}^{2}}=\frac{\sigma_{0}^{2}}{\Omega}\sum_{i=1}^N\left(\frac{\Theta^{2}}{\sigma_{\theta}^{2}}\right)^{2},
\end{equation}
where $\sigma_{0}^{2}$ as the reference measurement variance, which can be interpreted as the number of effective pixels within a chosen galaxy angular size, $\sigma_{0}=\Theta^{2}_0/\sigma_{\theta}^{2}=N_{\text{pix}}\equiv1$, here set to 1 for simplicity. Note that this definition is analogous, but not identical, to the commonly used effective number density in weak lensing surveys~\cite{Albrecht:2006um,Chang:2013xja}, and therefore should not be confused with it.
Thus, the effective noise PSD can be expressed in terms of $\tilde{n}_{\text{eff}}$ as
\begin{equation}\label{eq:Sn_eff_neff}
    S_n^{\mathrm{eff}} \simeq \frac{30}{7}\frac{\Delta t}{\sigma_{0}^{2}\Omega\tilde{n}_{\text{eff}}} = \frac{30}{7}\frac{\Delta t}{\Omega\tilde{n}_{\text{eff}}}.
\end{equation}

We derive the effective number density $\tilde{n}_{\text{eff}}$ for a survey as a function of its AB magnitude threshold $m_t$ in Sec.~\ref{sec:depth_to_neff}.

For a monochromatic signal at frequency $f_{\mathrm{gw}}$, which does not evolve within the observation time $T$, the integral in Eq.~\eqref{eq:SNR2_total} simplifies to give
\begin{equation}
    \left(\frac{S}{N}\right)_{\mathrm{total}} \approx \sqrt{\frac{h_0T}{S_n^{\mathrm{eff}}}}.
\end{equation}

However, over long observation times, the signal frequency can drift across bands. To account for this, and to compare sources and detectors on equal footing, we use the \emph{characteristic strain} $h_c(f)$ for the GW amplitude and the \emph{characteristic noise} $h_n(f)$ for the detector sensitivity, defined as~\cite{Moore:2014lga}
\begin{equation}
    \left(\frac{S}{N}\right)^{2}=4\int_{0}^{\infty}\frac{\left|\tilde{h}\left(f\right)\right|^{2}}{S_{n}\left(f\right)}df\equiv\int_{0}^{\infty}\frac{h_{c}^{2}\left(f\right)}{h_{n}^{2}\left(f\right)}\frac{df}{f},
\end{equation}
with $h_c(f)\equiv 2f|\tilde{h}(f)|$.
For inspiralling binaries, the characteristic strain becomes~\cite{Moore:2014lga}\footnote{Note that our definition of $h_c(f)$ differs by a factor of $\sqrt{2}$ from that in Ref.~\cite{Moore:2014lga}, since we denote $h_0$ as the \emph{peak} amplitude and not the root-mean-square amplitude.}
\begin{equation}\label{eq:binary_hc}
    h_c(f) = \sqrt{\frac{f^{2}}{\dot{f}}}h_{0},
\end{equation}
where $h_0$ is the strain amplitude and $\dot{f}$ is the frequency temporal derivative, given in Eq.~\eqref{eq:h0} and~\eqref{eq:dfdt} respectively. This prefactor can be interpreted as the square root of the number of cycles spent at frequency $f$, such that the characteristic strain of a GW from an inspiraling binary system scales as $h_c\propto f^{-1/6}$.

To keep the definition for $h_c(f)$ consistent across different GW detectors, we define the effective noise for a weak lensing survey with $N$ galaxies as
\begin{equation}\label{eq:hn}
    h_n(f) \equiv \sqrt{f S_n^{\mathrm{eff}}}.
\end{equation}

We note that for a survey with finite observation time $T$ and cadence $\Delta t$, the noise PSD $S_n(f)$ (and so $S_n^{\mathrm{eff}}$) is only defined over the frequency band $f\in[1/T,1/(2\Delta t)]$. Outside this band, we set $S_n(f)\rightarrow\infty$, which effectively truncates the sensitivity curves $h_n(f)$ at low and high frequencies. Additionally, we limit the characteristic strain $h_c(f)$ to frequencies below the innermost stable circular orbit (ISCO) of the binary system, given by $f_{\mathrm{ISCO}}=c^{3}/(6^{3/2}\pi GM_{t})$, where $M_t$ is the total mass of the binary.

%%%%%%%%%%%%%%%%%%%%%%%%%%%%%%%%%%%%%%%%%%%%%%%%%%%%%%%%%%
\subsection{Mapping Survey Depth to Effective number density}\label{sec:depth_to_neff}
%%%%%%%%%%%%%%%%%%%%%%%%%%%%%%%%%%%%%%%%%%%%%%%%%%%%%%%%%%

\begin{figure}[t]
    \centering
    \includegraphics[width=1\columnwidth]{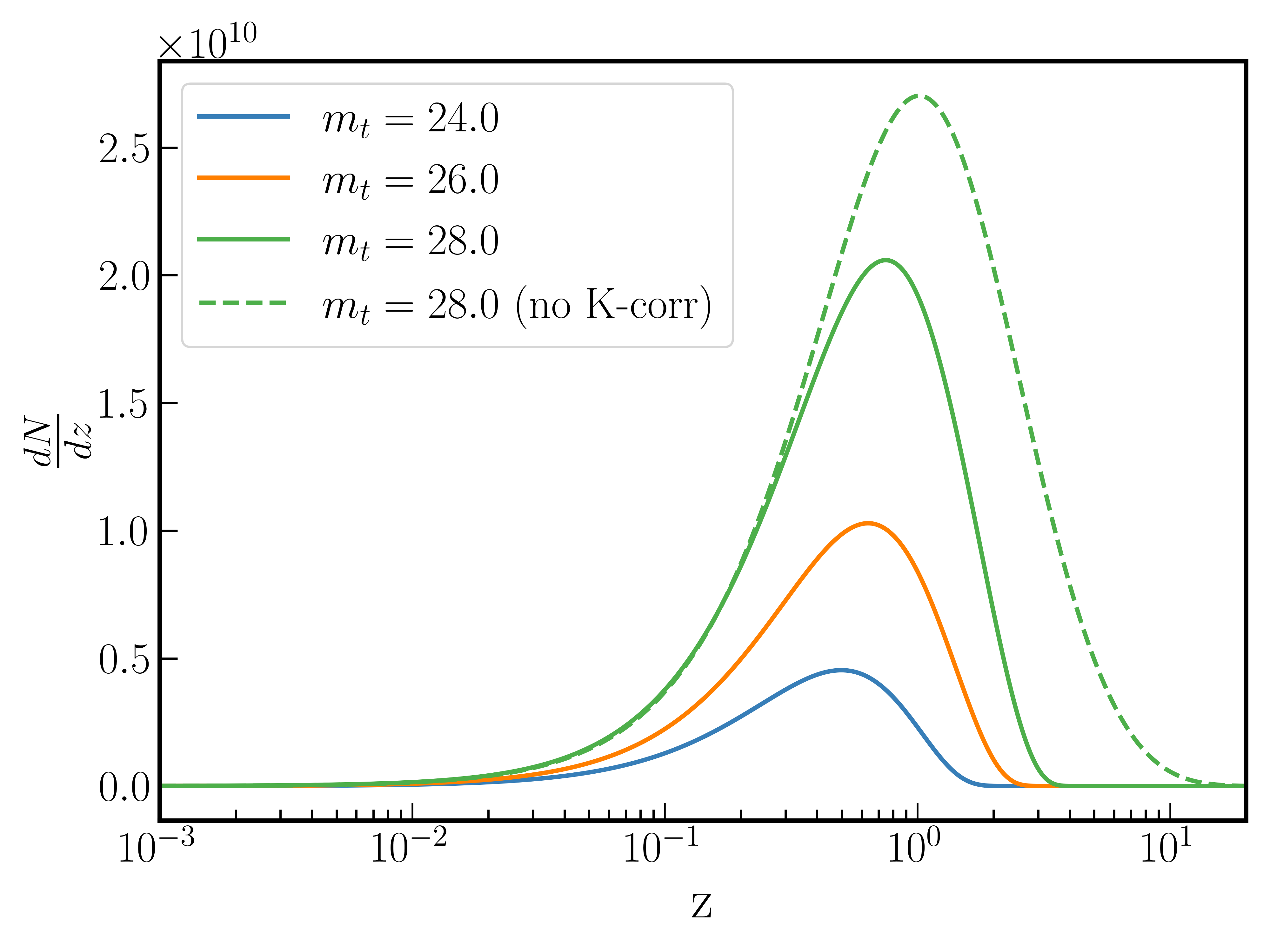}
    \caption{Redshift distribution of galaxies for different r-band AB magnitude thresholds $m_t$.}
    \label{fig:dndz_z}
\end{figure}

We relate the survey depth directly to the effective number density $\tilde{n}_{\text{eff}}$ defined in Eq.~\eqref{eq:n_eff_def} by modeling the galaxy population as a function of redshift and absolute magnitude. For a survey with solid angle $\Omega$ and effective angular resolution $\sigma_\theta$, this can be written as
\begin{align}
    \tilde n_\mathrm{eff} &= \frac{1}{\Omega}\int\frac{\Theta^{4}}{\sigma_{\theta}^{4}} dV_{c}\,\phi\left(z,M\right)\,dM\label{eq:n_eff_integral}\\
    &= \frac{1}{\sigma_{\theta}^{4}} \int dz\,\frac{c\chi^2(z)}{H(z)}\int_{-\infty}^{M_t(z)} dM\,\Theta^4\,\phi\left(M\right), \nonumber
\end{align}
where $dV_c=c\chi^2/H(z) dz d\Omega$ is the comoving volume element~\cite{Hogg:1999ad}, written in terms of the comoving distance $\chi(z)$ and the Hubble parameter $H(z)$. We assume a flat $\Lambda$CDM cosmology,
\begin{equation}
    H(z) = H_0\sqrt{\Omega_m(1+z)^3 + \Omega_\Lambda},
\end{equation}
with $H_0=70\,\text{km}\,\text{s}^{-1}\,\text{Mpc}^{-1}$, $\Omega_m=0.3$, and $\Omega_\Lambda=0.7$, and
\begin{equation}
    \chi(z) = c\int_0^z \frac{dz'}{H(z')}.
\end{equation}
The luminosity function, $\phi(z,M)$, determines the number density of galaxies per unit absolute magnitude $M$ at redshift $z$, and $M_t(z)$ denotes the absolute magnitude threshold corresponding to the survey's apparent magnitude threshold $m_t$ at redshift $z$, related through the distance modulus,
\begin{equation}
    M_t(z) = m_t - 5\log_{10}\left(\frac{d_L(z)}{10\;\mathrm{Mpc}}\right) - 25 - K(z).
\end{equation}
Here, $d_{L}\left(z\right)=\left(1+z\right)\chi\left(z\right)$ is the luminosity distance~\cite{Hogg:1999ad}, and $K(z)$ is the $K$-correction for the frequency-band shift in the observed spectral energy distribution due to redshift. For simplicity, we assume no redshift dependence in the luminosity function, and adopt a Schechter form~\cite{Schechter:1976iz},
\begin{align}
    \phi\left(M\right)dM &= 0.4\log\left(10\right)\phi_{*}10^{0.4\left(M_{*}-M\right)\left(\alpha+1\right)} \nonumber\\
    &\quad\times \exp\left[-10^{\frac{2}{5}\left(M_{*}-M\right)}\right]dM,\label{eq:phi_Schechter}
\end{align}
with best-fit r-band values for the parameters $\phi_{*}$, $M_{*}$, and $\alpha$ from Ref.~\cite{Loveday:2011dh}, and linear fit to the median $K$-corrections from the same reference\footnote{While we do not consider redshift dependence in the parameters, we confirmed that our results are robust to within 10\% when varying the parameters by 15\% around the fiducial values.}. We show the resulting redshift distribution of galaxies for different AB magnitude thresholds in Fig.~\ref{fig:dndz_z}, along with the distribution without $K$-correction for comparison. Here, $dN/dz=\Omega c\chi^2/H\int dM\,\phi(M)$, which is closely related to Eq.~\eqref{eq:n_eff_integral}.
We note that our modeling underestimates the expected number of galaxies for Vera Rubin Observatory's LSST and Euclid surveys~\cite{Euclid:2024yrr,LSST:2008ijt} by a factor of $\gtrsim2$, and can be therefore considered conservative.

We model the angular size of a galaxy appearing in Eq.~\eqref{eq:n_eff_integral} using
\begin{equation}
    \Theta\left(M,z\right)\simeq\frac{2r\left(M\right)}{d_{A}\left(z\right)},
\end{equation}
where $d_A(z)=\chi(z)/(1+z)$ is the angular diameter distance~\cite{Hogg:1999ad}, and $r(M)$ is the effective half-light radius of a galaxy with absolute magnitude $M$. To relate galaxy size to luminosity, we adopt a log-linear size-luminosity relation calibrated on early-type galaxies from using SDSS data from Ref.~\cite{Shen:2003sda}\footnote{While this relation was derived for low-redshift galaxies ($z\lesssim0.3$), we verified that our results are robust to within a factor of 2 when varying the parameters by 50\%, and in agreement with other reports~\cite{Cameron:2007sx,Newton:2011jw}.} (see Appendix~\ref{app:mag-size_relation}),
\begin{equation}
    r\left(M\right)=r_0 10^{kM},
\end{equation}
with $\log(r_0/\mathrm{kpc})=-5.535$ and $k=-0.143$.

Applying this model to Eq.~\eqref{eq:n_eff_integral}, the effective number density for a given survey configuration becomes
\begin{equation}\label{eq:n_eff_expression}
    \tilde n_\mathrm{eff}=\frac{8}{\sigma_{\theta}^{4}} \int_{0}^{\infty}dz\,\tilde{\phi}_{*}\frac{c\left(1+z\right)^{4}}{H\left(z\right)\chi^{2}\left(z\right)}\Gamma\left(\tilde{\alpha}+1,x_{t}\right), 
\end{equation}
where we defined
\begin{align}
    x_t&\equiv10^{\frac{2}{5}\left(M_{*}-M_{t}\right)} \nonumber\\
    \tilde{\alpha}&\equiv\alpha-10k,\\
    \tilde{\phi}_{*}&\equiv\phi_{*}r_{0}^{4}10^{4kM_{*}}.\nonumber
\end{align}

We note that the expression in the integral in Eq.~\eqref{eq:n_eff_expression} diverges at $z\rightarrow0$ due to the $\chi^{-2}(z)$ factor. This is expected, since the comoving volume scales as $\chi^3$, while $\Theta^4\propto\chi^{-4}$. To avoid this unphysical behavior, truncating the integral at a minimum redshift is not enough, as the increasing angular size of luminous galaxies bias the measurement error towards lower values. We treat this problem by adding a lower limit on the absolute magnitude integral in Eq.~\eqref{eq:n_eff_integral}, corresponding to the luminosity at which we expect only 1 galaxy within a comoving volume at each redshift, $M_{N=1}$, such that in practice
\begin{equation}\label{eq:gamma_truncation}
    \Gamma\left(\tilde{\alpha}+1,x_{t}\right) \rightarrow 
    \begin{cases}
        \Gamma_t - \Gamma_{N=1}, & x_t > x_{N=1}, \\
        0, & x_t \leq x_{N=1},
    \end{cases}
\end{equation}
where $\Gamma_i\equiv\Gamma\left(\tilde{\alpha}+1,x_{i}\right)$. We detail this procedure in Appendix~\ref{app:galaxy_size_snr}.

%%%%%%%%%%%%%%%%%%%%%%%%%%%%%%%%%%%%%%%%%%%%%%%%%%%%%%%%%%
\subsection{SGWB sensitivity}\label{sec:gwb_snr}
%%%%%%%%%%%%%%%%%%%%%%%%%%%%%%%%%%%%%%%%%%%%%%%%%%%%%%%%%%

Using Eq.~\eqref{eq:FX}-\eqref{eq:FX_rot}, Ref.~\cite{Mentasti:2024fgt} derived the angular correlations of the distortion components induced by a SGWB. The resulting SNR was given in Eq.~(46) of Ref.~\cite{Mentasti:2024fgt} as
\begin{align}
    \mathrm{SNR}_\kappa^2 &\simeq \left\langle \left|\Gamma^{\kappa\kappa}\left(\beta\right)\right|^{2}\right\rangle \frac{N^{2}T}{\sigma_{\kappa}^{2}\Delta t^{2}}\int_{1/T}^{1/\Delta t}df\,\left(\frac{h_{c}^{2}\left(f\right)}{16\pi f}\right)^{2} \nonumber\\
    &\simeq \left(\frac{N}{\sigma_{\kappa}^{2}\Delta t}\right)^2\frac{3}{288\pi^{4}}\frac{h_{\text{ref}}^{4}}{f_{\text{ref}}^{4\gamma}}\frac{T^{2-4\gamma}}{1-4\gamma},\label{eq:snr2_conv_gwb}
\end{align}
where $\Gamma^{\kappa\kappa}(\beta)$ is the angular correlation pattern for the convergence distortion component, $\beta$ is the angular separation between galaxy pairs (see Ref.~\cite{Mentasti:2024fgt}), and $h_c(f)=h_{\text{ref}}(f/f_{\text{ref}})^\gamma$ is the assumed characteristic strain spectrum. Here we consider the same reference values as in Ref.~\cite{Mentasti:2024fgt,NANOGrav:2023hfp,EPTA:2023xxk}, $h_{\text{ref}}=3\times 10^{-15}$ at $f_{\text{ref}}=3\times 10^{-8}$ Hz and $\gamma=-2/3$. Summing over all 9 possible correlators, the total SNR is given by $\mathrm{SNR}_\mathrm{total} \simeq 3\,\mathrm{SNR}_\kappa$.

\begin{figure}[t]
    \centering
    \includegraphics[width=1\columnwidth]{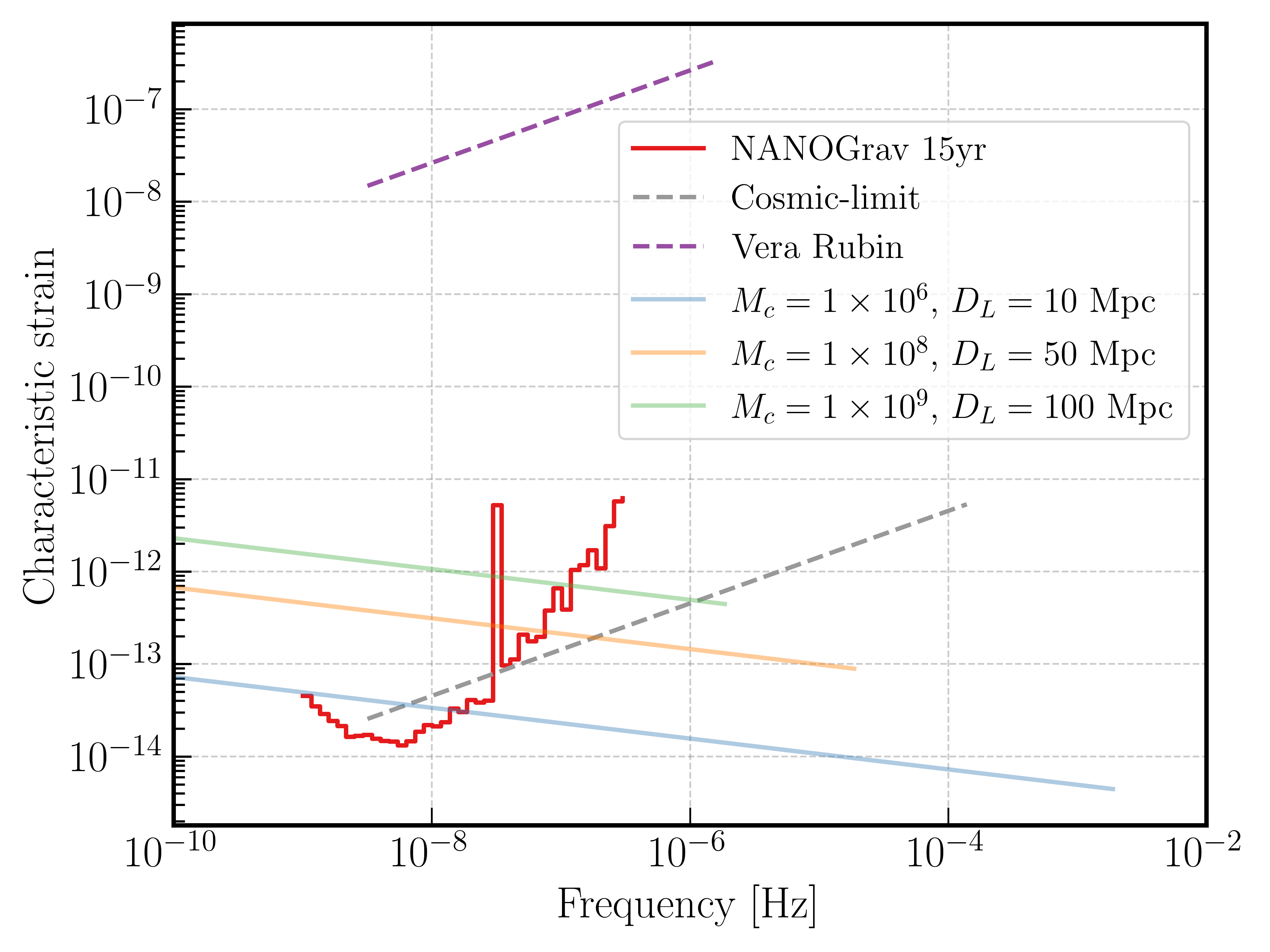}
    \caption{The characteristic noise $h_n(f)$ is shown for two weak-lensing survey configurations: (i) an LSST-like survey (purple dashed line) and (ii) a cosmic-limit survey (gray dashed line). For comparison, we also plot the effective sensitivity of NANOGrav 15yr~\cite{NANOGrav:2023pdq}, using $h_n=h_0\sqrt{f T_\mathrm{15yr}}$ with $T_\mathrm{15yr}=15$ years. Straight solid lines show the characteristic strain $h_c(f)$, in Eq.~\eqref{eq:binary_hc}, for circular inspiraling binaries with various chirp masses $\mathcal{M}_c$ and luminosity distances $d_L$ (see App.~\ref{app:waveform}), truncated at their ISCO frequencies.}
    \label{fig:hc_vs_f}
\end{figure}

The term in parenthesis in Eq.~\eqref{eq:snr2_conv_gwb} arises from the summation over all galaxy pairs with different noise PSD $S_{n,i}\simeq 2\Delta t\,\sigma_{\kappa,i}^2$, which, for large number of galaxies, yields
\begin{equation}
    \sum_{i<j}\frac{1}{\sigma_{\kappa,i}^{2}}\frac{1}{\sigma_{\kappa,j}^{2}}\simeq\frac{1}{2}\left(\sum_{i=1}^{N}\frac{1}{\sigma_{\kappa}^{2}}\right)^{2}.
\end{equation}
Assuming a fixed $\sigma_\kappa^2$ for all galaxies, leads to the factor $\left(N/\sigma_\kappa^2\Delta t\right)^2$ in Eq.~\eqref{eq:snr2_conv_gwb}. However, applying our treatment of the effective number density from Eq.~\eqref{eq:n_eff_def}, we can rewrite the \emph{total} SNR for the SGWB as
\begin{equation}\label{eq:SGWB_SNR}
    \mathrm{SNR}_\mathrm{total} ^2 \simeq \frac{27}{288\pi^{4}} \left(\frac{\Omega\,\tilde{n}_{\text{eff}}}{\Delta t}\right)^2\frac{h_{\text{ref}}^{4}}{f_{\text{ref}}^{4\gamma}}\frac{T^{2-4\gamma}}{1-4\gamma}.
\end{equation}

%%%%%%%%%%%%%%%%%%%%%%%%%%%%%%%%%%%%%%%%%%%%%%%%%%%%%%%%%%%
\section{Results and Discussion}\label{sec:results}
%%%%%%%%%%%%%%%%%%%%%%%%%%%%%%%%%%%%%%%%%%%%%%%%%%%%%%%%%%%

We analyze the sensitivity of weak-lensing surveys to GWs from inspiraling SMBHBs, quantified by the characteristic noise $h_n(f)$ defined in Eq.~\eqref{eq:hn}. The effective noise PSD, $S_n^{\mathrm{eff}}$, entering $h_n(f)$ is computed using Eq.~\eqref{eq:Sn_eff_neff} and~\eqref{eq:n_eff_expression}, which depends on the survey parameters: sky coverage $\Omega$, angular resolution $\sigma_\theta$, cadence $\Delta t$, and AB magnitude threshold $m_t$. The frequency band of the sensitivity curve is determined by the observation time $f_{\rm min}\approx 1/T$ and cadence $f_{\rm max}\approx 1/2\Delta t$.

Our analysis consists of two representative survey configurations: (i) Vera Rubin Observatory's LSST~\cite{LSST:2008ijt}, covering 18,000 deg$^2$, with r-band AB magnitude threshold $m_t=24.7$ (for single exposures)\footnote{We don't use the higher coadded sensitivity, as our signal is time dependent.}, angular resolution $\sigma_\theta=0.2\;\text{arcsec}$, observation time $T=10$ years, and cadence $\Delta t=3.5$ days; and (ii) a full-sky hypothetical survey with $m_t=28$, $\sigma_\theta=1\;\text{mas}$, $T=10$ years, and $\Delta t=1$ hour, representing the ``cosmic-limit'' for such surveys.

Our main results are shown in Fig.~\ref{fig:hc_vs_f}, depicting the sensitivity curves for the two survey configurations, along with the NANOGrav 15yr effective sensitivity curve~\cite{NANOGrav:2023pdq} for comparison. For indication, we also plot the characteristic strain $h_c(f)$ for a selection of circular inspiraling binaries with various chirp masses $\mathcal{M}_c$, located at different luminosity distances $d_L$, using the conventional formalism summarized in Appendix~\ref{app:waveform}.

\begin{figure}
    \centering
    \includegraphics[width=1\columnwidth]{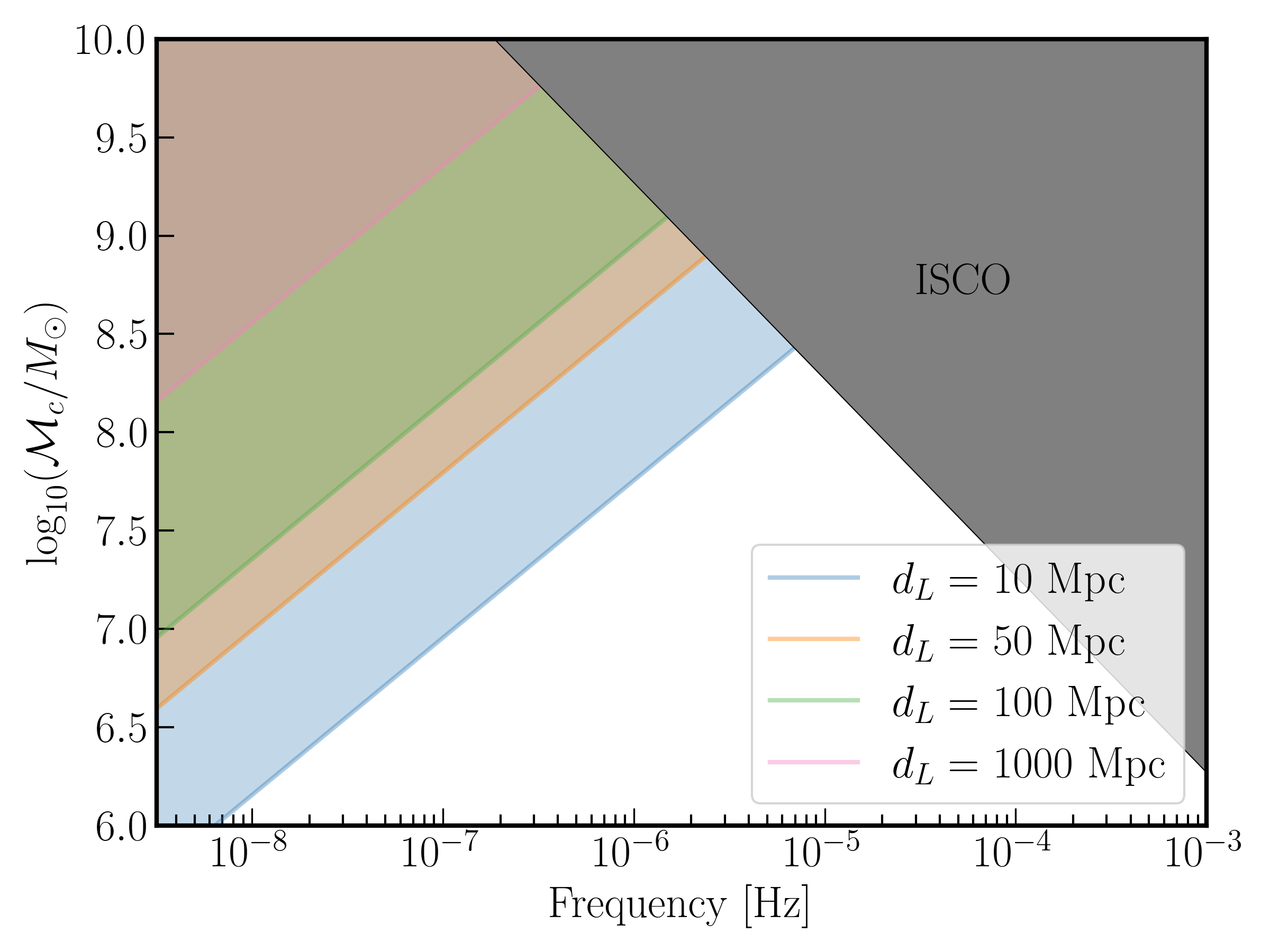}
    \caption{The detectable region in the chirp mass $\mathcal{M}_c$ and frequency $f$ parameter space for inspiraling binaries at different luminosity distances $d_L$, assuming the cosmic-limit survey configuration. The gray shaded region indicates frequencies above the ISCO for equal-mass binaries.}
    \label{fig:mc_vs_f}
\end{figure}

We find that LSST-like surveys are not sufficiently sensitive to detect GWs from inspiraling SMBHBs across realistic mass and distance ranges. Our modeling predicts that single exposures will yield roughly $\sim2.6$ billion detected galaxies. Although LSST's high cadence and long observation baseline enable it to probe GW frequencies in the range $f\sim10^{-8}-10^{-6}$ Hz—a band that overlaps with PTAs and partially complements future space-based detectors such as LISA—its limited angular resolution leads to a large effective noise level. As a result, the survey remains insensitive to the expected SMBHB signals despite its extensive sky coverage and temporal sampling.
In contrast, the cosmic-limit configuration—with over 30~billion detected galaxies—exhibits a significantly improved sensitivity, becoming competitive with PTAs in the $\sim 10$ nHz band and partially bridging the frequency gap between PTAs and LISA. To quantify its reach, we map the detectable region in chirp mass $\mathcal{M}_c$ and luminosity distance $d_L$, defined by the condition $h_c = h_n$ across the accessible frequency range, as shown in Fig.~\ref{fig:mc_vs_f}. We find that such a survey could detect binaries with $\mathcal{M}_c \gtrsim 10^8 M_\odot$ out to luminosity distances of several Gpc, depending on the frequency.

This comparison highlights the large gap—spanning several orders of magnitude—between the sensitivity of realistic surveys such as LSST and the idealized cosmic-limit case. Bridging this gap would require advances in angular resolution and survey design that go well beyond currently foreseeable capabilities.
\begin{figure}
    \centering
    \includegraphics[width=1\columnwidth]{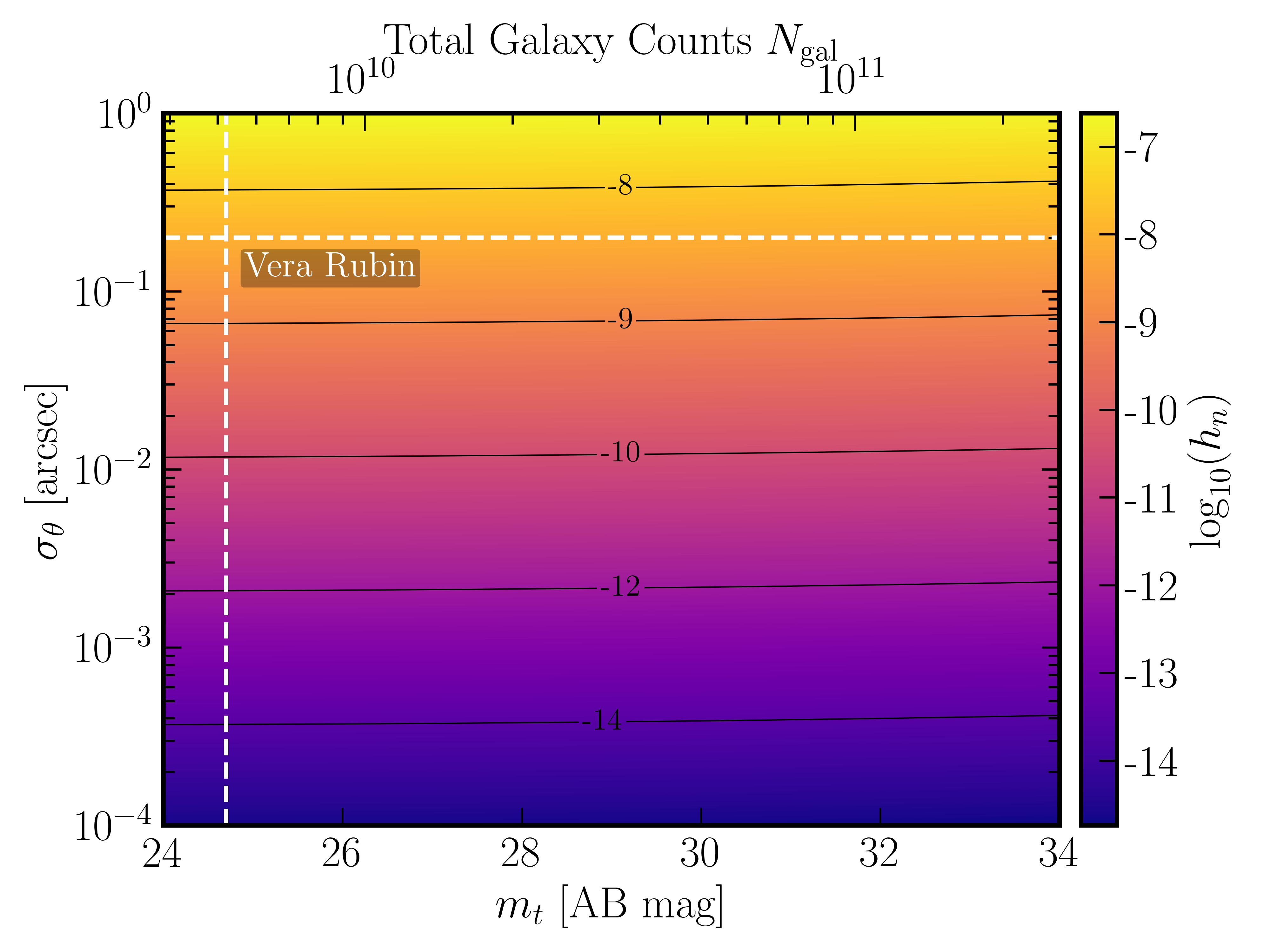}
    \caption{Contours of the characteristic noise $h_n$ at $f=10$ nHz as a function of the survey r-band AB magnitude threshold $m_t$ and angular resolution $\sigma_\theta$, assuming full-sky coverage and cadence $\Delta t=1$ day. The top axis shows the corresponding total number of observed galaxies $N_\mathrm{gal}$ in the survey. White dashed lines indicate the parameter values of Vera Rubin's LSST.}
    \label{fig:mt_vs_res}
\end{figure}

Next, we investigate how the survey parameters affect the characteristic noise. In Fig.~\ref{fig:mt_vs_res}, we plot contours of the characteristic noise at a fixed frequency, $f=10$ nHz, as a function of the effective AB magnitude threshold $m_t$ and angular resolution $\sigma_\theta$. In this plot, we considered a more conservative cadence of $\Delta t=1$ day, with full-sky coverage. We find, as expected from the scaling in Eq.~\eqref{eq:hn}-\eqref{eq:Sn_sigma}, $ h_n \propto \sigma_\theta^2$, that the angular resolution is the dominant factor in improving sensitivity. Furthermore, while deepening the survey (increasing $m_t$) significantly boosts the number of observed galaxies, the corresponding improvement in sensitivity is negligible. This behavior can be understood from the galaxy and angular-size distributions. Increasing the limiting magnitude $m_t$ primarily adds intrinsically faint galaxies to the sample. These galaxies are both physically smaller and, in principle, detectable out to higher redshifts. However, the latter gain is strongly suppressed by the K-correction, which rapidly diminishes the visibility of distant galaxies in the optical bands, as shown in Figure~\ref{fig:dndz_z}. As a result, deeper imaging essentially adds smaller galaxies, whose contribution to the effective signal is weighted by $\Theta^4$ and is therefore minor. The sensitivity is thus dominated by the relatively few large, low-redshift galaxies, which are already included at modest values of $m_t$.

\begin{figure}
    \centering
    \includegraphics[width=1\columnwidth]{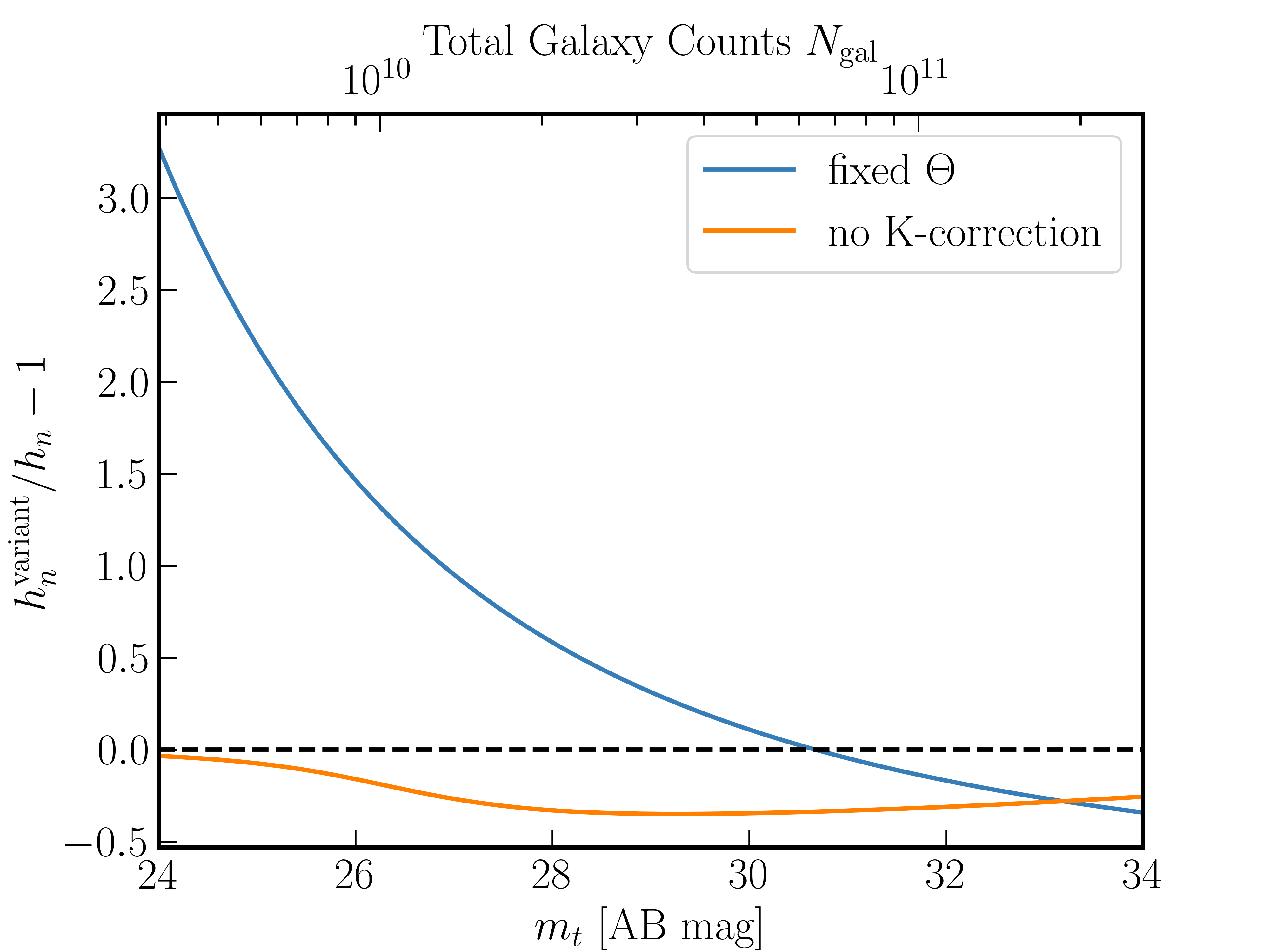}
    \caption{The relative difference in $h_n$, with the baseline given by the $\sigma_\theta = 0.01$ arcsec slice in Fig.~\ref{fig:mt_vs_res}, is shown for two modified scenarios: (i) fixing all galaxy angular sizes to $\Theta=0.7$ arcsec (blue), and (ii) omitting the K-correction in the distance modulus (red).}
    \label{fig:hn_comparison}
\end{figure}

Extending the observable wavelength range so that galaxies can be detected efficiently at $z \gtrsim 2$ would change this picture. Beyond this redshift, the angular diameter distance begins to decrease, causing galaxies to subtend larger apparent sizes and therefore contribute more strongly to the signal. This expectation is confirmed in Figure~\ref{fig:hn_comparison}. When the angular sizes are artificially held fixed, $h_n$ is enhanced by a factor of $\sim 3$ at low $m_t$, but then decreases and falls below the baseline beyond $m_t\simeq 31$. Conversely, removing the K-correction leaves $h_n$ unchanged at low depths but leads to a steep improvement at higher $m_t$, saturating at a lower asymptotic value. This behavior mirrors the trends in Figure~\ref{fig:dndz_z}, where the redshift distribution widens substantially once the K-correction is omitted, increasing the relative weight of high-redshift, large-angular-size galaxies.

Finally, using our more careful treatment of $\tilde n_\mathrm{eff}$—in contrast to the fixed-galaxy assumptions adopted in earlier work~\cite{Mentasti:2024fgt}—we obtain the following numerical estimate for the SGWB SNR corresponding to Eq.~\eqref{eq:SGWB_SNR}:
\begin{align}
    \mathrm{SNR}_\mathrm{total} &\simeq 0.02\left(\frac{T}{10\;\text{years}}\right)^{7/3}\left(\frac{\text{hour}}{\Delta t}\right)\\
    &\quad\times\left(\frac{\text{mas}}{\sigma_{\theta}}\right)^{4}\left(\frac{m_{t}}{28}\right)^{0.9}.\nonumber
\end{align}
Here we approximated the scaling of $\tilde{n}_{\text{eff}}$ with $m_t$ as $\tilde{n}_{\text{eff}}\approx 7.35\times10^9\left(\mathrm{mas}/\sigma_\theta\right)^4 m_t^{0.9}\;\left[\mathrm{arcsec^{-2}}\right]$, which holds for $m_t\in[22,32]$ within 5\% accuracy. This expression shows that detecting the SGWB with weak-lensing surveys is even more challenging than that of continuous waves, relying on extremely high angular resolution ($\sigma_\theta \lesssim 0.1$ mas) to reach an SNR of order unity over decade-long observations.

For completeness, we note that Doppler-tracking experiments such as Cassini probe a similar frequency range~\cite{Tinto:1998ee,Abbate:2003stu,Armstrong:2003ay}, but existing results are expressed as constraints on a SGWB rather than on coherent CW sources, and are therefore not shown in Figure~\ref{fig:hc_vs_f}. A direct comparison would require a dedicated CW reanalysis of the Doppler data, which is beyond the scope of this work.

While the cosmic-variance-limited sensitivity derived here provides a useful theoretical benchmark, achieving such performance would require observational capabilities far beyond those of current or planned surveys. The benchmark shown in Figure~\ref{fig:hc_vs_f} assumes full-sky coverage; reducing the survey area by a factor of a few would degrade the sensitivity only at the level of an order-unity correction, and therefore does not qualitatively change this conclusion. Instead, the dominant challenges arise from the simultaneous requirements of mas angular resolution, wide-area coverage, and high-cadence (hourly) observations. For example, diffraction-limited imaging at $\sim$mas resolution in the optical would require apertures of order tens to hundreds of meters in space, together with a survey strategy capable of repeatedly imaging a large fraction of the sky at high depth on hour timescales. By comparison, current wide-field surveys such as the Vera Rubin Observatory achieve sub-arcsecond resolution with cadences of days, while space-based missions like Gaia~\cite{Gaia:2016zol} reach much higher astrometric precision but are not designed for time-domain weak-lensing measurements of resolved galaxy shapes.
In addition, systematic effects such as time-dependent point-spread function variations, detector anisotropies, and calibration drifts (e.g., small changes in pixel geometry or optical distortions over time) would need to be controlled at extremely stringent levels. These requirements, particularly on angular resolution, cadence, and instrument stability over the relevant timescales, underscore that the cosmic-variance-limited case should be interpreted as an idealized upper bound rather than a realizable near-term experimental target.
%%%%%%%%%%%%%%%%%%%%%%%%%%%%%%%%%%%%%%%%%%%%%%%%%%%%%%%%%%%
\section{Conclusion}\label{sec:conclusion}
%%%%%%%%%%%%%%%%%%%%%%%%%%%%%%%%%%%%%%%%%%%%%%%%%%%%%%%%%%%

Gravitational waves (GWs) offer a new and exciting avenue for exploring the Universe. As the LIGO-Virgo-KAGRA collaboration continues to detect GWs from stellar-mass binary mergers, and pulsar timing arrays (PTAs) report evidence for a stochastic GW background (SGWB) at nanohertz frequencies, there is growing interest in expanding GW observations to other frequency bands. The nanohertz to microhertz range is particularly intriguing, as it is expected to be populated by signals from supermassive black hole binaries (SMBHBs) and could bridge the gap between PTA and space-based GW detectors like LISA. In this work, we have investigated the potential of weak-lensing surveys to detect continuous waves (CWs) in this frequency range by measuring the induced distortions in the shapes of distant galaxies. Building on previous theoretical frameworks, we have developed a more realistic model that accounts for the distribution of galaxy sizes and redshifts, as well as survey parameters such as sky coverage, angular resolution, cadence, and depth. Our analysis reveals, as shown in Figure~\ref{fig:hc_vs_f}, that current and near-future weak-lensing surveys, such as LSST, are unlikely to detect CWs from inspiraling SMBHBs due to their limited angular resolution. A hypothetical cosmic-limit survey, with angular resolution of $\sim \mathrm{mas}$ and hourly cadence, could achieve competitive sensitivity in the nanohertz band and extend it to higher frequencies. Such a survey could detect SMBHBs with chirp masses $\mathcal{M}_c \gtrsim 10^8 M_\odot$ out to several Gpc, partially bridging the frequency gap between PTAs and LISA. We also find that the angular resolution of the survey is the most critical factor in improving sensitivity, while increasing depth has a relatively minor effect due to the dominance of nearby, large-angular-size galaxies in the signal. Extending the observable wavelength range to detect galaxies at higher redshifts could enhance sensitivity by incorporating larger apparent sizes. However, extending our analysis to higher redshift surveys will involve additional factors such as intergalactic medium attenuation~\cite{Meiksin:2005gw}, improved modeling of galaxy redshift distribution~\cite{Tortorelli:2024ogu}, and calibration of the size-depth scaling. Finally, we estimate that detecting the SGWB with weak-lensing surveys would require extremely high angular resolution ($\sigma_\theta \lesssim 0.1$ mas) to achieve a signal-to-noise ratio of order unity over decade-long observations, making such a detection extremely challenging in practice. That said, cross-correlating time-domain shear measurements with relative astrometric deflection or PTA time residuals probing the same frequency band~\cite{Mentasti:2025ttn,Vaglio:2025tex,Fink:2025wlk} could help suppress survey-specific systematics and enhance detection robustness, making weak-lensing measurements a complementary probe alongside PTA or astrometric data. Overall, our results highlight both the challenges and opportunities in using weak-lensing surveys for GW detection. While such surveys are unlikely to be competitive as standalone probes of low-frequency GWs in the nHz–$\mu$Hz range with current technology, future advancements in observational capabilities—particularly when combined with multi-tracer cross-correlations—could open new windows into the low-frequency GW universe.

%%%%%%%%%%%%%%%%%%%%%%%%%%%%%%%%%%%%%%%%%%%%%%%%%%%%%%%%%%%

\begin{acknowledgments}
    We thank the anonymous referee for constructive comments that helped improve the clarity of this work. We thank Priyank Parashari and José Benavides for the useful discussions. TA is supported in part by the Zuckerman STEM Leadership Program. Part of this work was done at Jet Propulsion Laboratory, California Institute of Technology, under a contract with the National Aeronautics and Space Administration (80NM0018D0004).
\end{acknowledgments}

%%%%%%%%%%%%%%%%%%%%%%%%%%%%%%%%%%%%%%%%%%%%%%%%%%%%%%%%%%%
\appendix
%%%%%%%%%%%%%%%%%%%%%%%%%%%%%%%%%%%%%%%%%%%%%%%%%%%%%%%%%%%

%%%%%%%%%%%%%%%%%%%%%%%%%%%%%%%%%%%%%%%%%%%%%%%%%%%%%%%%%%%
\section{Waveform and Polarization Conventions}\label{app:waveform}
%%%%%%%%%%%%%%%%%%%%%%%%%%%%%%%%%%%%%%%%%%%%%%%%%%%%%%%%%%%%

The metric perturbation associated with a plane GW propagating along the unit direction $\hat{q}$ can be expressed in terms of its two linear polarizations, $h_{+}(t)$ and $h_{\times}(t)$, as
\begin{equation}
    h_{ij}\left(t,\hat{q}\right) = 
        h_{+}\left(t\right)e_{ij}^{\left(+\right)}\left(\hat{q}\right)
        + h_{\times}\left(t\right)e_{ij}^{\left(\times\right)}\left(\hat{q}\right),
\end{equation}
where,
\begin{align}
    e_{ij}^{\left(+\right)}\left(\hat{q}\right) &= \hat{u}_i\hat{u}_j - \hat{v}_i\hat{v}_j, \\
    e_{ij}^{\left(\times\right)}\left(\hat{q}\right) &= \hat{u}_i\hat{v}_j + \hat{v}_i\hat{u}_j, \nonumber
\end{align}
with $\hat{u}$ and $\hat{v}$ being two orthonormal vectors transverse to $\hat{q}$. Alternatively, the strain can be expressed in terms of the left and right circular polarizations, by defining
\begin{align}
    h_{L}\left(t\right) &= \frac{1}{2}\left[h_{+}(t) - i\,h_{\times}(t)\right], \label{eq:circular_polarizations}\\
    h_{R}\left(t\right) &= \frac{1}{2}\left[h_{+}(t) + i\,h_{\times}(t)\right], \nonumber
\end{align}
and the corresponding polarization tensors
\begin{align}
    e_{ij}^{\left(L\right)}\left(\hat{q}\right) &= e_{ij}^{\left(+\right)} + i\,e_{ij}^{\left(\times\right)}, \\
    e_{ij}^{\left(R\right)}\left(\hat{q}\right) &= e_{ij}^{\left(+\right)} - i\,e_{ij}^{\left(\times\right)}. \nonumber
\end{align}
This form can be generalized by introducing an arbitrary polarization angle $\psi$, which corresponds to a rotation of the basis vectors $\hat{u}$ and $\hat{v}$ about the direction $\hat{q}$, adding a phase factor to the circular polarization components, i.e.,
\begin{equation}
    h_{L}\rightarrow h_{L}e^{-2i\psi},\quad\text{ and }\quad h_{R}\rightarrow h_{R}e^{+2i\psi}.
\end{equation}

The linear polarization amplitudes for a GW from an inspiraling binary system in a circular orbit with orbital frequency $f_{\text{orb}}$ are given by~\cite{Maggiore:2007ulw}
\begin{align}
    h_{+}\left(t\right)&=h_{0}\frac{1+\cos^{2}\iota}{2}\cos\left[\Phi(t)\right], \label{eq:linear_polarizations}\\
    h_{\times}\left(t\right)&=h_{0}\cos\iota\sin\left[\Phi(t)\right], \nonumber
\end{align}
where $\iota$ is the inclination angle of the binary orbital plane with respect to the line of sight, and $h_0$ is the strain amplitude given by
\begin{equation}\label{eq:h0}
    h_{0}=\frac{4\left(G\mathcal{M}_{c}\right)^{5/3}\left(\pi f_{\text{gw}}\right)^{2/3}}{c^{4}d_{L}}
\end{equation}
Here, $G$ is the gravitational constant, $c$ the speed of light, $d_L$ the luminosity distance to the source, and $\mathcal{M}_c$ is the chirp mass of the binary, defined as
\begin{equation}
    \mathcal{M}_c = \frac{(m_1 m_2)^{3/5}}{(m_1 + m_2)^{1/5}}.
\end{equation}
The phase evolution $\Phi(t)$ can be approximated to have a constant frequency over short observation times,
\begin{equation}
    \Phi(t) = 2\pi f_{\text{gw}} t + \phi_0,
\end{equation}
with the GW frequency evolving as
\begin{equation}\label{eq:dfdt}
    \dot{f}_{\text{gw}}=\frac{96}{5}\pi^{8/3}\left(\frac{G\mathcal{M}_{c}}{c^{3}}\right)^{5/3}f_{\text{gw}}^{11/3}.
\end{equation}

%%%%%%%%%%%%%%%%%%%%%%%%%%%%%%%%%%%%%%%%%%%%%%%%%%%%%%%%%%%
\section{Derivation of the Distortion Tensor}\label{app:gw_dist}
%%%%%%%%%%%%%%%%%%%%%%%%%%%%%%%%%%%%%%%%%%%%%%%%%%%%%%%%%%%

In this appendix, we closely follow the derivation in Ref.~\cite{Mentasti:2024fgt} to derive the expression for the response functions $F_X^p(\hat{q},\hat{n})$ in Eq.~\eqref{eq:FX}.
Starting from the expression for the deflection angle induced by a passing GW in 
Eq.~\eqref{eq:deflection}, we compute the distortion tensor by taking the angular gradient with respect to the source direction $\hat{n}$,
\begin{align}
    \tilde{\psi}_{ij}\left(\hat{q},\hat{n}\right)&\equiv\frac{\partial\delta\hat{n}_{i}}{\partial\hat{n}^{j}} \\
    &= \frac{1}{2}\left[\delta_{ij}-\frac{\hat{n}_{i}+\hat{q}_{i}}{1+\hat{q}_{l}\hat{n}^{l}}\hat{q}_{j}\right]\frac{h_{rk}\hat{n}^{r}\hat{n}^{k}}{1+\hat{q}_{l}\hat{n}^{l}} \nonumber \\
    &\quad +\frac{\hat{n}_{i}+\hat{q}_{i}}{1+\hat{q}_{l}\hat{n}^{l}}h_{jk}\hat{n}^{k}-\frac{1}{2}h_{ij} . \nonumber
\end{align}
We then use spherical polar coordinates centered at the observer's location to write down $\hat{q}$ and $\hat{n}$,
\begin{align}
    \hat{q}&=\cos\phi\sin\theta\,\hat{x}+\sin\phi\sin\theta\,\hat{y}+\cos\theta\,\hat{z},\\
    \hat{n}&=\cos\phi_{s}\sin\theta_{s}\,\hat{x}+\sin\phi_{s}\sin\theta_{s}\,\hat{y}+\cos\theta_{s}\,\hat{z}, \nonumber
\end{align}
and recover the tangential distortion matrix by projecting onto the local basis vectors, as defined in Eq.~\eqref{eq:psi_ab}, with
\begin{align}
    \hat{e}_{\theta}\left(\hat{n}\right)&=\left(\cos\phi_{s}\cos\theta_{s},\,\sin\phi_{s}\cos\theta_{s},\,-\sin\theta_{s}\right),\\
    \hat{e}_{\phi}\left(\hat{n}\right)&=\left(-\sin\phi_{s},\,\cos\phi_{s},0\right).\nonumber
\end{align}

We then use the same convention introduced in Appendix~\ref{app:waveform} to write the left-circular polarization tensor in terms of the orthogonal basis vectors, $\bs{v}^L(\hat q)\equiv \hat e_\theta(\hat q) + i \hat e_\phi(\hat q)$, as
\begin{equation}
    h_{ij}^{L}\left(\hat{q}\right)=\left[\boldsymbol{v}^{L}\otimes\boldsymbol{v}^{L}\right]_{ij}\left(\hat{q}\right)=v_{i}^{L}\left(\hat{q}\right)v_{j}^{L}\left(\hat{q}\right).
\end{equation}
After some algebra, we find that the \emph{left-circular} distortion tensor can be expressed as
\begin{align}
    \tilde{\psi}_{ij}^{L}\left(\hat{q},\hat{n}\right) &= \frac{1}{2}\left[\delta_{ij}-\frac{\hat{n}_{i}+\hat{q}_{i}}{1+\hat{q}\cdot\hat{n}}\hat{q}_{j}\right]\frac{\left[\hat{n}\cdot\boldsymbol{v}^{L}\left(\hat{q}\right)\right]^{2}}{1+\hat{q}\cdot\hat{n}} \nonumber \\
    &\quad +\frac{\hat{n}_{i}+\hat{q}_{i}}{1+\hat{q}\cdot\hat{n}}\left[\hat{n}\cdot\boldsymbol{v}^{L}\left(\hat{q}\right)\right]v_{j}^{L}\left(\hat{q}\right) \\
    &\quad -\frac{1}{2}v_{i}^{L}\left(\hat{q}\right)v_{j}^{L}\left(\hat{q}\right). \nonumber
\end{align}
Next, we choose the local basis vectors for the tangential projection as\footnote{We note that there is a typo in Eq.~(17) of Ref.~\cite{Mentasti:2024fgt}, where the definitions of $\hat{e}_\theta$ is orthogonal to $\hat q$ instead of $\hat n$.}
\begin{align}
    \hat{e}_{\phi}\left(\hat{q},\hat{n}\right)&=\frac{\hat{n}\times\hat{q}}{\sqrt{1-\left(\hat{n}\cdot\hat{q}\right)^{2}}},\\
    \hat{e}_{\theta}\left(\hat{q},\hat{n}\right)&=\frac{\left(\hat{n}\times\hat{q}\right)\times\hat{n}}{\sqrt{1-\left(\hat{n}\cdot\hat{q}\right)^{2}}},
\end{align}
such that $\hat{e}_{\theta}\times\hat{e}_{\phi}=\hat{n}$. 

Projecting onto this basis, we find that the distortion matrix~\eqref{eq:psi_ab} can be expressed as
\begin{widetext}
    \begin{align}
        \tilde{\psi}_{ab}^{L}\left(\hat{q},\hat{n}\right) &= \hat{e}_{a}^{i}\left(\hat{q},\hat{n}\right)\tilde{\psi}_{ij}^{L}\left(\hat{q},\hat{n}\right)\hat{e}_{b}^{j}\left(\hat{q},\hat{n}\right) \label{eq:psyL_ab_nq}\\
        &= \frac{1}{2}\frac{\left[\hat{n}\cdot\boldsymbol{v}^{L}\left(\hat{q}\right)\right]^{2}}{1+\hat{q}\cdot\hat{n}}\left[\delta_{ab}-\frac{\left[\hat{e}_{a}\left(\hat{q},\hat{n}\right)\cdot\hat{q}\right]\left[\hat{e}_{b}\left(\hat{q},\hat{n}\right)\cdot\hat{q}\right]}{1+\hat{q}\cdot\hat{n}}\right]+\frac{\hat{n}\cdot\boldsymbol{v}^{L}\left(\hat{q}\right)}{1+\hat{q}\cdot\hat{n}}\left[\hat{e}_{a}\left(\hat{q},\hat{n}\right)\cdot\hat{q}\right]\left[\hat{e}_{b}\left(\hat{q},\hat{n}\right)\cdot\boldsymbol{v}^{L}\right] \nonumber \\
        &\quad -\frac{1}{2}\left[\hat{e}_{a}\left(\hat{q},\hat{n}\right)\cdot\boldsymbol{v}^{L}\left(\hat{q}\right)\right]\left[\hat{e}_{b}\left(\hat{q},\hat{n}\right)\cdot\boldsymbol{v}^{L}\left(\hat{q}\right)\right], \nonumber
    \end{align}
\end{widetext}
which can be recast in the form of Eq.~\eqref{eq:psi_ab_decomp}, such that
\begin{equation}
    \tilde{\psi}_{ab}^{L}\left(\hat{q},\hat{n}\right)\equiv\left(\begin{array}{cc}
-F_{\kappa}-F_{\gamma_{1}} & -F_{\gamma_{2}}+F_{\omega}\\
-F_{\gamma_{2}}-F_{\omega} & -F_{\kappa}+F_{\gamma_{1}}
\end{array}\right).
\end{equation}
For the special case where the GW is propagating along the $\hat{z}$-direction, i.e., $\hat{q}=\hat{z}$, our choice of basis vectors yields
\begin{align}
    \bs{v}^{L}\left(\hat{z}\right)&=e^{-i\phi}\left(\hat{x}+i\hat{y}\right) , \\
    \hat{e}_{\phi}\left(\hat{z},\hat{n}\right) &= \sin\phi_{s}\hat{x}-\cos\phi_{s}\hat{y},\nonumber\\
    \hat{e}_{\theta}\left(\hat{z},\hat{n}\right) &= -\cos\phi_{s}\cos\theta_{s}\hat{x}-\sin\phi_{s}\cos\theta_{s}\hat{y}+\sin\theta_{s}\hat{z}. \nonumber
\end{align}
Plugging these into Eq.~\eqref{eq:psyL_ab_nq}, and doing some algebra, we arrive at
\begin{align}
    \tilde{\psi}_{ab}^{L}\left(\hat{z},\hat{n}\right) &= \frac{1}{2}e^{-2i\left(\phi-\phi_{s}\right)} \\
    &\quad\times\left(
        \begin{array}{cc}
            -\cos\theta_{s} & i\cos\theta_{s}-2i\\
            -i\cos\theta_{s} & 2-\cos\theta_{s}
        \end{array}\right), \nonumber
\end{align}
where we identify the response functions for the left-circular polarization as
\begin{align}
    F_{\kappa}^{L}\left(\hat{z},\hat{n}\right) &= -\frac{1}{2}e^{-2i\left(\phi-\phi_{s}\right)}\left(1-\cos\theta_{s}\right),\nonumber\\
    F_{\omega}^{L}\left(\hat{z},\hat{n}\right) &= -iF_{\kappa}^{L}\left(\hat{z},\hat{n}\right),\\
    F_{\gamma_{1}}^{L}\left(\hat{z},\hat{n}\right) &= \frac{1}{2}e^{-2i\left(\phi-\phi_{s}\right)},\nonumber \\
    F_{\gamma_{2}}^{L}\left(\hat{z},\hat{n}\right) &= -iF_{\gamma_{1}}^{L}\left(\hat{z},\hat{n}\right).\nonumber
\end{align}
Using Eq.~\eqref{eq:complex_shear} we get to the final expression for the left-circular shear response function in Eq.~\eqref{eq:FX}.

%%%%%%%%%%%%%%%%%%%%%%%%%%%%%%%%%%%%%%%%%%%%%%%%%%%%%%%%%%%
\section{Matched-Filter SNR: Detailed Derivation}\label{app:matched_filter}
%%%%%%%%%%%%%%%%%%%%%%%%%%%%%%%%%%%%%%%%%%%%%%%%%%%%%%%%%%%

In this appendix, we provide a detailed derivation of the matched-filter SNR described in Sec.~\ref{sec:snr_formalism}.
We begin by defining the Fourier transform
\begin{align}
    \mathcal{F}\left[\cos\Phi\right]&\equiv\frac{1}{2}\int_{-\infty}^{\infty}dt\left(e^{\Phi-2\pi ift}+e^{-\left(\Phi+2\pi ift\right)}\right)\\
    &\equiv\Delta_{+}+\Delta_{-},\nonumber
\end{align}
such that the Fourier transform of the strain amplitudes in Eq.~\eqref{eq:linear_polarizations} can be expressed as
\begin{align}
    \tilde{h}_{+}\left(f\right)&=\frac{h_{0}\left(1+\cos^{2}\iota\right)}{2}\left(\Delta_{+}+\Delta_{-}\right), \\
    \tilde{h}_{\times}\left(f\right)&=\frac{h_{0}\cos\iota}{i}\left(\Delta_{+}-\Delta_{-}\right). \nonumber
\end{align}
The corresponding circular polarization amplitudes are then given by
\begin{align}
    \tilde{h}_{L}\left(f\right)&=\frac{h_{0}}{4}\left[\left(1-\cos\iota\right)^{2}\Delta_{+}+\left(1+\cos\iota\right)^{2}\Delta_{-}\right], \\
    \tilde{h}_{R}\left(f\right)&=\frac{h_{0}}{4}\left[\left(1+\cos\iota\right)^{2}\Delta_{+}+\left(1-\cos\iota\right)^{2}\Delta_{-}\right]. \nonumber
\end{align}
We then consider Eq.~\eqref{eq:SNR2}, averaging over the polarization angle $\psi$ and inclination angle $\iota$. The average over $\psi$ removes cross-terms between left and right polarizations, which eliminates the cross terms in the squared modulus, yielding
\begin{equation}
    \left\langle \left(\frac{S_{X}}{N_{X}}\right)^{2}\right\rangle _{\psi} = 4\int_{0}^{\infty}\frac{\sum_{p}\left|F_{X}^{p}\left(\hat{q},\hat{n}\right)\tilde{h}_{p}\left(f\right)\right|^{2}}{S_{n}^{X}\left(f\right)}\,df
\end{equation}
Next, we average over the inclination angle $\iota$, noting that
\begin{align}
    \frac{1}{2}\int_{0}^{\pi}&\left[\left(1-\cos\iota\right)^{2}\Delta_{+}+\left(1+\cos\iota\right)^{2}\Delta_{-}\right]^{2}\sin\iota\,d\iota \nonumber \\
    &=\frac{16}{5}\left|\Delta_{+}+\Delta_{-}\right|^{2},
\end{align}
and similarly for the right polarization. The resulting expression for the numerator becomes
\begin{align}
    \sum_{p}\left|F_{X}^{p}\left(\hat{q},\hat{n}\right)\tilde{h}_{p}\left(f\right)\right|^{2}&=\frac{2}{5}h_{0}^{2}\left|\Delta_{+}+\Delta_{-}\right|^{2}\\
    &\times\left|F_{X}^{L}\left(\hat{q},\hat{n}\right)\right|^{2},
\end{align}
where we used the symmetry between left and right polarizations, i.e., $\left|F_{X}^{R}\right|^{2}=\left|F_{X}^{L}\right|^{2}$. Plugging this back into the SNR expression yields
\begin{equation}
    \left\langle \left(\frac{S_{X}}{N_{X}}\right)^{2}\right\rangle _{\psi,\iota} = \frac{8}{5}\left|F_{X}^{L}\left(\hat{q},\hat{n}\right)\right|^{2}\int_{0}^{\infty}\frac{\left|\mathcal{F}\left[h_{0}\cos\Phi\right]\right|^{2}}{S_{n}^{X}\left(f\right)}\,df, 
\end{equation}
which is Eq.~\eqref{eq:SNR2_final} in the main text.

\begin{figure}
    \centering
    \includegraphics[width=1\columnwidth]{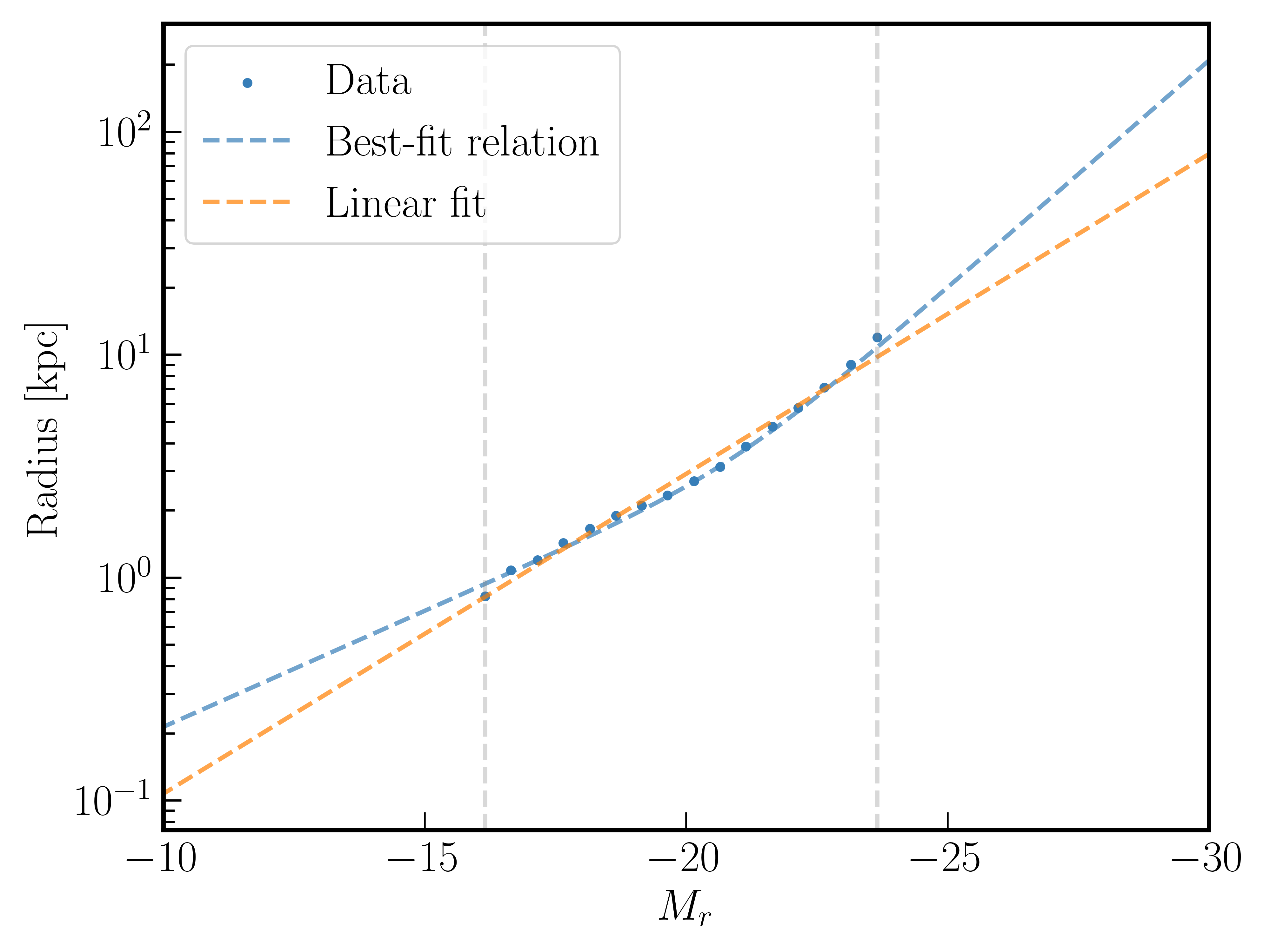}
    \caption{Effective half-light radius $r$ as a function of absolute magnitude $M$ in the r-band for early-type galaxies from Ref.~\cite{Shen:2003sda} (blue points). The blue dashed line corresponds to the best-fit model in Ref.~\cite{Shen:2003sda}, while the orange dashed line shows our linear fit model.}
    \label{fig:mag_to_r}
\end{figure}

The last step is to average over the full sky. Using the response functions in Eq.~\eqref{eq:FX}, and their rotation properties in Eq.~\eqref{eq:FX_rot}, it's easy to show that
\begin{align}
    \left\langle \left|F_{X}^{L}\left(\hat{q},\hat{n}\right)\right|^{2}\right\rangle_\Omega &= \frac{1}{4\pi}\int d\Omega_{\hat{n}'}\left|F_{X}^{L}\left(\hat{z},\hat{n}'\right)\right|^{2} \\
    &= \begin{cases}
        \frac{1}{3}, & X=\kappa,\omega, \\
        1, & X={}_{+2}\gamma,
    \end{cases} \nonumber
\end{align}
so that summing over all distortion components gives
\begin{equation}
    \sum_{X}\left\langle \left|F_{X}^{L}\left(\hat{q},\hat{n}\right)\right|^{2}\right\rangle_\Omega = \frac{7}{3}.
\end{equation}
Plugging this back into the SNR expression yields Eq.~\eqref{eq:SNR2_avg} in the main text.

%%%%%%%%%%%%%%%%%%%%%%%%%%%%%%%%%%%%%%%%%%%%%%%%%%%%%%%%%%%
\section{Magnitude-Size Relation}\label{app:mag-size_relation}
%%%%%%%%%%%%%%%%%%%%%%%%%%%%%%%%%%%%%%%%%%%%%%%%%%%%%%%%%%%

To relate the galaxy absolute magnitude $M$ to its physical size $r$, we use the empirical relation shown in Ref.~\cite{Shen:2003sda} for early-type galaxies observed in the Sloan Digital Sky Survey (SDSS). We fit the data from Figure~6 in Ref.~\cite{Shen:2003sda} to a linear relation in log-space, relating the effective half-light radius $r$ (in kpc) scales with the absolute magnitude $M$ (in the r-band).
As shown in Figure~\ref{fig:mag_to_r}, our simplified model underestimates the sizes of both bright and faint galaxies, but is mostly consistent with the typical sizes for intermediate magnitudes, leading to a conservative estimate of the effective number density.

%%%%%%%%%%%%%%%%%%%%%%%%%%%%%%%%%%%%%%%%%%%%%%%%%%%%%%%%%%%
\section{Treatment of Galaxy Size in the SNR Estimator}\label{app:galaxy_size_snr}
%%%%%%%%%%%%%%%%%%%%%%%%%%%%%%%%%%%%%%%%%%%%%%%%%%%%%%%%%%%

\begin{figure}[t]
    \centering
    \includegraphics[width=1\columnwidth]{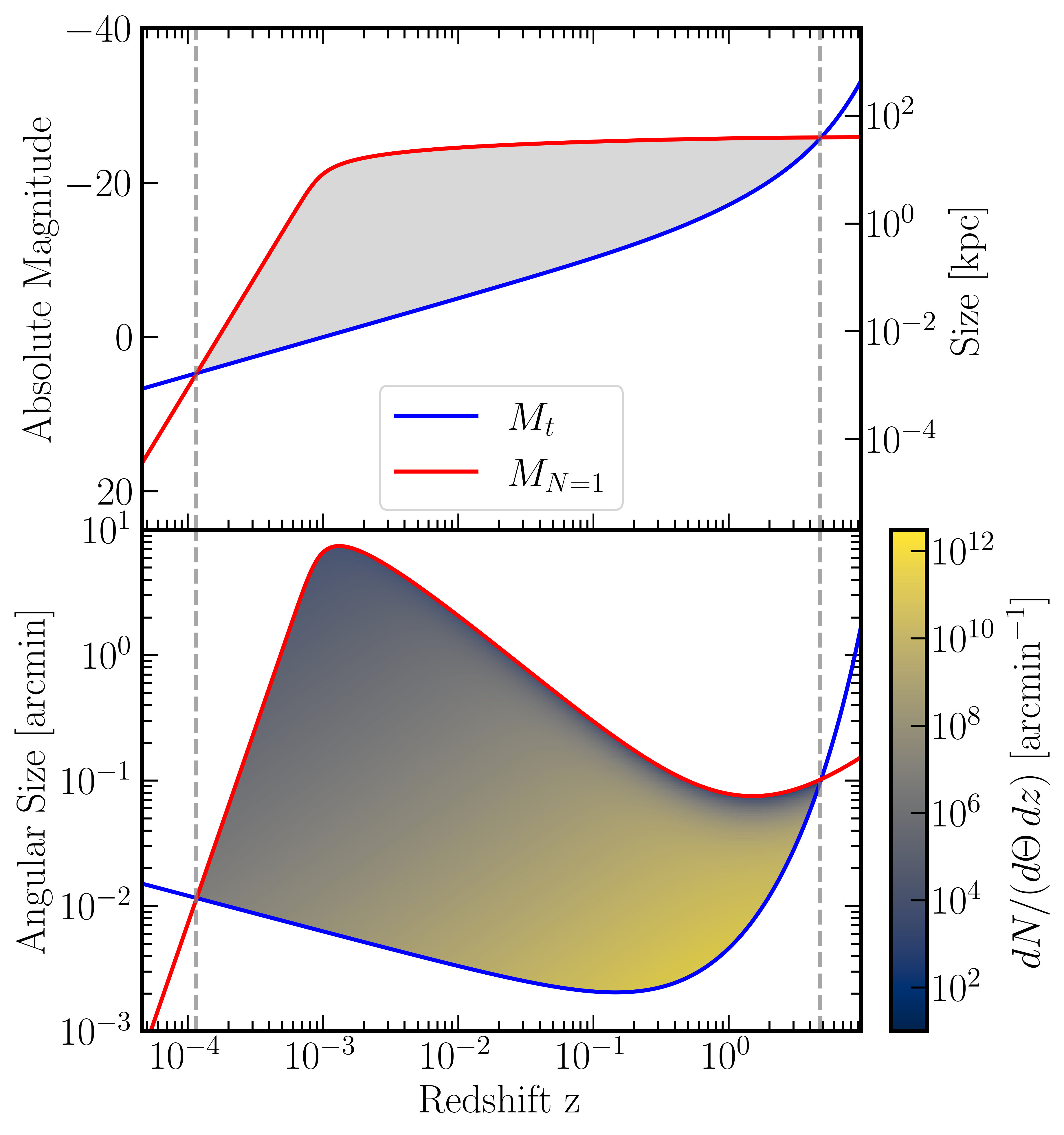}
    \caption{Absolute magnitude limits $M_{N=1}$ (red) and $M_t$ (blue) as a function of redshift (top panel), together with their corresponding angular sizes (bottom panel). The results shown assume a full-sky configuration with $m_t=28$. In the bottom panel, the colormap shows the differential number of galaxies per unit angular size and redshift, $dN/(d\Theta\,dz)$, while the shaded region indicates the integration domain in Eq.~\eqref{eq:n_eff_integral}. The vertical dashed lines mark the redshifts at which the two magnitude limits intersect.}
    \label{fig:size_vs_z}
\end{figure}

Here we detail our treatment of the galaxy angular size $\Theta$ in the effective number density in Eq.~\eqref{eq:n_eff_integral}. As discussed in Sec.~\ref{sec:depth_to_neff}, the angular size increases at low redshifts, leading to unrealistically small measurement errors. To avoid this problem, we set a lower limit on the absolute magnitude integral in Eq.~\eqref{eq:n_eff_integral}, corresponding to the absolute magnitude at which we expect only 1 galaxy within a comoving volume at each redshift, $M_{N=1}$, defined implicitly as
\begin{equation}
    1=\Omega\int_{0}^{z}dz'\;\frac{c\chi^{2}\left(z'\right)}{H\left(z'\right)}\phi\left(z',M_{N=1}\right).
\end{equation}
Since we assume a redshift independent luminosity function, we can integrate this expression to obtain
\begin{equation}
    \phi\left(M_{N=1}\right)=\frac{3}{\Omega\chi^{3}(z)},
\end{equation}
where we used $d\chi/dz=c/H\left(z\right)$. For the Schechter luminosity function in Eq.~\eqref{eq:phi_Schechter}, this condition can be solved analytically for $M_{N=1}$ as follows:
\begin{align}
    M_{N=1} &= M_{*}-\frac{5}{2}\frac{\log\left(x_{N=1}\right)}{\log\left(10\right)}\\
    &=M_{*}-\frac{5}{2}\frac{1}{\log\left(10\right)}\log\left[ -\left(1+\alpha\right)W\left(z\right)\right],
\end{align}
where $x_{N=1}=10^{0.4\left(M_{*}-M_{N=1}\right)}$ and $W(z)$ is the Lambert W function, defined as the solution to $w e^{w}=z$~\cite{Corless:1996zz}, with
\begin{equation}
    z = -\frac{\left(\frac{5\phi}{\phi_*\log\left(100\right)}\right)^{\frac{1}{1+\alpha}}}{1+\alpha}.
\end{equation}
Therefore, in practice, we implement the following replacement in Eq.~\eqref{eq:n_eff_integral}:
\begin{widetext}
\begin{equation}
    \int_{-\infty}^{M_t} dM \rightarrow \int_{M_{N=1}}^{M_t} dM =
    \begin{cases}
        \int_{-\infty}^{M_t} dM - \int_{-\infty}^{M_{N=1}} dM, & M_t > M_{N=1}, \\
        0, & M_t \leq M_{N=1},
    \end{cases}
\end{equation}
\end{widetext}
leading to Eq.~\eqref{eq:gamma_truncation}.

In Figure~\ref{fig:size_vs_z} we plot the resulting magnitude limits $M_{N=1}$ and $M_t$, along with their corresponding galaxy physical and angular sizes as a function of redshift. We see how the maximum angular size is bounded and remains below $\sim10$ arcmin, and crosses the minimum angular size set by $M_t$ at $z\sim10^{-4}$ and angular size of $\sim0.06$ arcsec. The crossing points set the effective redshift cutoffs in the SNR integral in Eq.~\eqref{eq:n_eff_integral}, where there are no galaxies bright enough to satisfy both the magnitude and number density limits.

\bibliography{refs.bib}

%merlin.mbs apsrev4-1.bst 2010-07-25 4.21a (PWD, AO, DPC) hacked
%Control: key (0)
%Control: author (0) dotless jnrlst
%Control: editor formatted (1) identically to author
%Control: production of article title (0) allowed
%Control: page (1) range
%Control: year (0) verbatim
%Control: production of eprint (0) enabled
\begin{thebibliography}{70}%
\makeatletter
\providecommand \@ifxundefined [1]{%
 \@ifx{#1\undefined}
}%
\providecommand \@ifnum [1]{%
 \ifnum #1\expandafter \@firstoftwo
 \else \expandafter \@secondoftwo
 \fi
}%
\providecommand \@ifx [1]{%
 \ifx #1\expandafter \@firstoftwo
 \else \expandafter \@secondoftwo
 \fi
}%
\providecommand \natexlab [1]{#1}%
\providecommand \enquote  [1]{``#1''}%
\providecommand \bibnamefont  [1]{#1}%
\providecommand \bibfnamefont [1]{#1}%
\providecommand \citenamefont [1]{#1}%
\providecommand \href@noop [0]{\@secondoftwo}%
\providecommand \href [0]{\begingroup \@sanitize@url \@href}%
\providecommand \@href[1]{\@@startlink{#1}\@@href}%
\providecommand \@@href[1]{\endgroup#1\@@endlink}%
\providecommand \@sanitize@url [0]{\catcode `\\12\catcode `\$12\catcode `\&12\catcode `\#12\catcode `\^12\catcode `\_12\catcode `\%12\relax}%
\providecommand \@@startlink[1]{}%
\providecommand \@@endlink[0]{}%
\providecommand \url  [0]{\begingroup\@sanitize@url \@url }%
\providecommand \@url [1]{\endgroup\@href {#1}{\urlprefix }}%
\providecommand \urlprefix  [0]{URL }%
\providecommand \Eprint [0]{\href }%
\providecommand \doibase [0]{http://dx.doi.org/}%
\providecommand \selectlanguage [0]{\@gobble}%
\providecommand \bibinfo  [0]{\@secondoftwo}%
\providecommand \bibfield  [0]{\@secondoftwo}%
\providecommand \translation [1]{[#1]}%
\providecommand \BibitemOpen [0]{}%
\providecommand \bibitemStop [0]{}%
\providecommand \bibitemNoStop [0]{.\EOS\space}%
\providecommand \EOS [0]{\spacefactor3000\relax}%
\providecommand \BibitemShut  [1]{\csname bibitem#1\endcsname}%
\let\auto@bib@innerbib\@empty
%</preamble>
\bibitem [{\citenamefont {Abbott}\ \emph {et~al.}(2016)\citenamefont {Abbott} \emph {et~al.}}]{LIGOScientific:2016aoc}%
  \BibitemOpen
  \bibfield  {author} {\bibinfo {author} {\bibfnamefont {B.~P.}\ \bibnamefont {Abbott}} \emph {et~al.} (\bibinfo {collaboration} {LIGO Scientific, Virgo}),\ }\bibfield  {title} {\enquote {\bibinfo {title} {{Observation of Gravitational Waves from a Binary Black Hole Merger}},}\ }\href {\doibase 10.1103/PhysRevLett.116.061102} {\bibfield  {journal} {\bibinfo  {journal} {Phys. Rev. Lett.}\ }\textbf {\bibinfo {volume} {116}},\ \bibinfo {pages} {061102} (\bibinfo {year} {2016})},\ \Eprint {http://arxiv.org/abs/1602.03837} {arXiv:1602.03837 [gr-qc]} \BibitemShut {NoStop}%
\bibitem [{\citenamefont {Maggiore}(2000)}]{Maggiore:1999vm}%
  \BibitemOpen
  \bibfield  {author} {\bibinfo {author} {\bibfnamefont {Michele}\ \bibnamefont {Maggiore}},\ }\bibfield  {title} {\enquote {\bibinfo {title} {{Gravitational wave experiments and early universe cosmology}},}\ }\href {\doibase 10.1016/S0370-1573(99)00102-7} {\bibfield  {journal} {\bibinfo  {journal} {Phys. Rept.}\ }\textbf {\bibinfo {volume} {331}},\ \bibinfo {pages} {283--367} (\bibinfo {year} {2000})},\ \Eprint {http://arxiv.org/abs/gr-qc/9909001} {arXiv:gr-qc/9909001} \BibitemShut {NoStop}%
\bibitem [{\citenamefont {Aasi}\ \emph {et~al.}(2015)\citenamefont {Aasi} \emph {et~al.}}]{LIGOScientific:2014pky}%
  \BibitemOpen
  \bibfield  {author} {\bibinfo {author} {\bibfnamefont {J.}~\bibnamefont {Aasi}} \emph {et~al.} (\bibinfo {collaboration} {LIGO Scientific}),\ }\bibfield  {title} {\enquote {\bibinfo {title} {{Advanced LIGO}},}\ }\href {\doibase 10.1088/0264-9381/32/7/074001} {\bibfield  {journal} {\bibinfo  {journal} {Class. Quant. Grav.}\ }\textbf {\bibinfo {volume} {32}},\ \bibinfo {pages} {074001} (\bibinfo {year} {2015})},\ \Eprint {http://arxiv.org/abs/1411.4547} {arXiv:1411.4547 [gr-qc]} \BibitemShut {NoStop}%
\bibitem [{\citenamefont {Acernese}\ \emph {et~al.}(2015)\citenamefont {Acernese} \emph {et~al.}}]{VIRGO:2014yos}%
  \BibitemOpen
  \bibfield  {author} {\bibinfo {author} {\bibfnamefont {F.}~\bibnamefont {Acernese}} \emph {et~al.} (\bibinfo {collaboration} {VIRGO}),\ }\bibfield  {title} {\enquote {\bibinfo {title} {{Advanced Virgo: a second-generation interferometric gravitational wave detector}},}\ }\href {\doibase 10.1088/0264-9381/32/2/024001} {\bibfield  {journal} {\bibinfo  {journal} {Class. Quant. Grav.}\ }\textbf {\bibinfo {volume} {32}},\ \bibinfo {pages} {024001} (\bibinfo {year} {2015})},\ \Eprint {http://arxiv.org/abs/1408.3978} {arXiv:1408.3978 [gr-qc]} \BibitemShut {NoStop}%
\bibitem [{\citenamefont {Akutsu}\ \emph {et~al.}(2021)\citenamefont {Akutsu} \emph {et~al.}}]{KAGRA:2020tym}%
  \BibitemOpen
  \bibfield  {author} {\bibinfo {author} {\bibfnamefont {T.}~\bibnamefont {Akutsu}} \emph {et~al.} (\bibinfo {collaboration} {KAGRA}),\ }\bibfield  {title} {\enquote {\bibinfo {title} {{Overview of KAGRA: Detector design and construction history}},}\ }\href {\doibase 10.1093/ptep/ptaa125} {\bibfield  {journal} {\bibinfo  {journal} {PTEP}\ }\textbf {\bibinfo {volume} {2021}},\ \bibinfo {pages} {05A101} (\bibinfo {year} {2021})},\ \Eprint {http://arxiv.org/abs/2005.05574} {arXiv:2005.05574 [physics.ins-det]} \BibitemShut {NoStop}%
\bibitem [{\citenamefont {Abac}\ \emph {et~al.}(2025{\natexlab{a}})\citenamefont {Abac}, \citenamefont {Abouelfettouh}, \citenamefont {Acernese} \emph {et~al.}}]{LIGOScientific:2025slb}%
  \BibitemOpen
  \bibfield  {author} {\bibinfo {author} {\bibfnamefont {A.G.}\ \bibnamefont {Abac}}, \bibinfo {author} {\bibfnamefont {I.}~\bibnamefont {Abouelfettouh}}, \bibinfo {author} {\bibfnamefont {F.}~\bibnamefont {Acernese}},  \emph {et~al.} (\bibinfo {collaboration} {LIGO Scientific}),\ }\bibfield  {title} {\enquote {\bibinfo {title} {{GWTC-4.0: Updating the Gravitational-Wave Transient Catalog with Observations from the First Part of the Fourth LIGO-Virgo-KAGRA Observing Run}},}\ }\href@noop {} {\  (\bibinfo {year} {2025}{\natexlab{a}})},\ \Eprint {http://arxiv.org/abs/2508.18082} {arXiv:2508.18082 [gr-qc]} \BibitemShut {NoStop}%
\bibitem [{\citenamefont {Abac}\ \emph {et~al.}(2025{\natexlab{b}})\citenamefont {Abac}, \citenamefont {Abouelfettouh}, \citenamefont {Acernese} \emph {et~al.}}]{LIGOScientific:2025pvj}%
  \BibitemOpen
  \bibfield  {author} {\bibinfo {author} {\bibfnamefont {A.G.}\ \bibnamefont {Abac}}, \bibinfo {author} {\bibfnamefont {I.}~\bibnamefont {Abouelfettouh}}, \bibinfo {author} {\bibfnamefont {F.}~\bibnamefont {Acernese}},  \emph {et~al.} (\bibinfo {collaboration} {LIGO Scientific}),\ }\bibfield  {title} {\enquote {\bibinfo {title} {{GWTC-4.0: Population Properties of Merging Compact Binaries}},}\ }\href@noop {} {\  (\bibinfo {year} {2025}{\natexlab{b}})},\ \Eprint {http://arxiv.org/abs/2508.18083} {arXiv:2508.18083 [astro-ph.HE]} \BibitemShut {NoStop}%
\bibitem [{\citenamefont {Abac}\ \emph {et~al.}(2025{\natexlab{c}})\citenamefont {Abac}, \citenamefont {Abouelfettouh}, \citenamefont {Acernese} \emph {et~al.}}]{LIGOScientific:2025rid}%
  \BibitemOpen
  \bibfield  {author} {\bibinfo {author} {\bibfnamefont {A.G.}\ \bibnamefont {Abac}}, \bibinfo {author} {\bibfnamefont {I.}~\bibnamefont {Abouelfettouh}}, \bibinfo {author} {\bibfnamefont {F.}~\bibnamefont {Acernese}},  \emph {et~al.} (\bibinfo {collaboration} {LIGO Scientific}),\ }\bibfield  {title} {\enquote {\bibinfo {title} {{GW250114: Testing Hawking’s Area Law and the Kerr Nature of Black Holes}},}\ }\href {\doibase 10.1103/kw5g-d732} {\bibfield  {journal} {\bibinfo  {journal} {Phys. Rev. Lett.}\ }\textbf {\bibinfo {volume} {135}},\ \bibinfo {pages} {111403} (\bibinfo {year} {2025}{\natexlab{c}})},\ \Eprint {http://arxiv.org/abs/2509.08054} {arXiv:2509.08054 [gr-qc]} \BibitemShut {NoStop}%
\bibitem [{\citenamefont {Abac}\ \emph {et~al.}(2025{\natexlab{d}})\citenamefont {Abac}, \citenamefont {Abouelfettouh}, \citenamefont {Acernese} \emph {et~al.}}]{LIGOScientific:2025brd}%
  \BibitemOpen
  \bibfield  {author} {\bibinfo {author} {\bibfnamefont {A.G.}\ \bibnamefont {Abac}}, \bibinfo {author} {\bibfnamefont {I.}~\bibnamefont {Abouelfettouh}}, \bibinfo {author} {\bibfnamefont {F.}~\bibnamefont {Acernese}},  \emph {et~al.} (\bibinfo {collaboration} {LIGO Scientific}),\ }\bibfield  {title} {\enquote {\bibinfo {title} {{GW241011 and GW241110: Exploring Binary Formation and Fundamental Physics with Asymmetric, High-spin Black Hole Coalescences}},}\ }\href {\doibase 10.3847/2041-8213/ae0d54} {\  (\bibinfo {year} {2025}{\natexlab{d}}),\ 10.3847/2041-8213/ae0d54},\ \Eprint {http://arxiv.org/abs/2510.26931} {arXiv:2510.26931 [astro-ph.HE]} \BibitemShut {NoStop}%
\bibitem [{\citenamefont {Amaro-Seoane}\ \emph {et~al.}(2017)\citenamefont {Amaro-Seoane}, \citenamefont {Audley}, \citenamefont {Babak} \emph {et~al.}}]{LISA:2017pwj}%
  \BibitemOpen
  \bibfield  {author} {\bibinfo {author} {\bibfnamefont {Pau}\ \bibnamefont {Amaro-Seoane}}, \bibinfo {author} {\bibfnamefont {Heather}\ \bibnamefont {Audley}}, \bibinfo {author} {\bibfnamefont {Stanislav}\ \bibnamefont {Babak}},  \emph {et~al.} (\bibinfo {collaboration} {LISA}),\ }\bibfield  {title} {\enquote {\bibinfo {title} {{Laser Interferometer Space Antenna}},}\ }\href@noop {} {\  (\bibinfo {year} {2017})},\ \Eprint {http://arxiv.org/abs/1702.00786} {arXiv:1702.00786 [astro-ph.IM]} \BibitemShut {NoStop}%
\bibitem [{\citenamefont {Luo}\ \emph {et~al.}(2015)\citenamefont {Luo}, \citenamefont {Chen}, \citenamefont {Duan} \emph {et~al.}}]{TianQin:2015yph}%
  \BibitemOpen
  \bibfield  {author} {\bibinfo {author} {\bibfnamefont {Jun}\ \bibnamefont {Luo}}, \bibinfo {author} {\bibfnamefont {Li-Sheng}\ \bibnamefont {Chen}}, \bibinfo {author} {\bibfnamefont {Hui-Zong}\ \bibnamefont {Duan}},  \emph {et~al.} (\bibinfo {collaboration} {TianQin}),\ }\bibfield  {title} {\enquote {\bibinfo {title} {{TianQin: a space-borne gravitational wave detector}},}\ }\href {\doibase 10.1088/0264-9381/33/3/035010} {\  (\bibinfo {year} {2015}),\ 10.1088/0264-9381/33/3/035010},\ \Eprint {http://arxiv.org/abs/1512.02076} {arXiv:1512.02076 [astro-ph.IM]} \BibitemShut {NoStop}%
\bibitem [{\citenamefont {Kawamura}\ \emph {et~al.}(2019)\citenamefont {Kawamura} \emph {et~al.}}]{Kawamura:2018esd}%
  \BibitemOpen
  \bibfield  {author} {\bibinfo {author} {\bibfnamefont {Seiji}\ \bibnamefont {Kawamura}} \emph {et~al.},\ }\bibfield  {title} {\enquote {\bibinfo {title} {{Space gravitational-wave antennas DECIGO and B-DECIGO}},}\ }\href {\doibase 10.1142/S0218271818450013} {\bibfield  {journal} {\bibinfo  {journal} {Int. J. Mod. Phys. D}\ }\textbf {\bibinfo {volume} {28}},\ \bibinfo {pages} {1845001} (\bibinfo {year} {2019})}\BibitemShut {NoStop}%
\bibitem [{\citenamefont {Baker}\ \emph {et~al.}(2019)\citenamefont {Baker}, \citenamefont {Bellovary}, \citenamefont {Bender} \emph {et~al.}}]{Baker:2019nia}%
  \BibitemOpen
  \bibfield  {author} {\bibinfo {author} {\bibfnamefont {John}\ \bibnamefont {Baker}}, \bibinfo {author} {\bibfnamefont {Jillian}\ \bibnamefont {Bellovary}}, \bibinfo {author} {\bibfnamefont {Peter~L.}\ \bibnamefont {Bender}},  \emph {et~al.},\ }\bibfield  {title} {\enquote {\bibinfo {title} {{The Laser Interferometer Space Antenna: Unveiling the Millihertz Gravitational Wave Sky}},}\ }\href@noop {} {\  (\bibinfo {year} {2019})},\ \Eprint {http://arxiv.org/abs/1907.06482} {arXiv:1907.06482 [astro-ph.IM]} \BibitemShut {NoStop}%
\bibitem [{\citenamefont {Antoniadis}\ \emph {et~al.}(2023{\natexlab{a}})\citenamefont {Antoniadis}, \citenamefont {Arumugam}, \citenamefont {Arumugam} \emph {et~al.}}]{EPTA:2023fyk}%
  \BibitemOpen
  \bibfield  {author} {\bibinfo {author} {\bibfnamefont {J.}~\bibnamefont {Antoniadis}}, \bibinfo {author} {\bibfnamefont {P.}~\bibnamefont {Arumugam}}, \bibinfo {author} {\bibfnamefont {S.}~\bibnamefont {Arumugam}},  \emph {et~al.} (\bibinfo {collaboration} {EPTA}),\ }\bibfield  {title} {\enquote {\bibinfo {title} {{The second data release from the European Pulsar Timing Array - III. Search for gravitational wave signals}},}\ }\href {\doibase 10.1051/0004-6361/202346844} {\bibfield  {journal} {\bibinfo  {journal} {Astron. Astrophys.}\ }\textbf {\bibinfo {volume} {678}},\ \bibinfo {pages} {A50} (\bibinfo {year} {2023}{\natexlab{a}})},\ \Eprint {http://arxiv.org/abs/2306.16214} {arXiv:2306.16214 [astro-ph.HE]} \BibitemShut {NoStop}%
\bibitem [{\citenamefont {Agazie}\ \emph {et~al.}(2023{\natexlab{a}})\citenamefont {Agazie}, \citenamefont {Anumarlapudi}, \citenamefont {Archibald} \emph {et~al.}}]{NANOGrav:2023gor}%
  \BibitemOpen
  \bibfield  {author} {\bibinfo {author} {\bibfnamefont {Gabriella}\ \bibnamefont {Agazie}}, \bibinfo {author} {\bibfnamefont {Akash}\ \bibnamefont {Anumarlapudi}}, \bibinfo {author} {\bibfnamefont {Anne~M.}\ \bibnamefont {Archibald}},  \emph {et~al.} (\bibinfo {collaboration} {NANOGrav}),\ }\bibfield  {title} {\enquote {\bibinfo {title} {{The NANOGrav 15 yr Data Set: Evidence for a Gravitational-wave Background}},}\ }\href {\doibase 10.3847/2041-8213/acdac6} {\  (\bibinfo {year} {2023}{\natexlab{a}}),\ 10.3847/2041-8213/acdac6},\ \Eprint {http://arxiv.org/abs/2306.16213} {arXiv:2306.16213 [astro-ph.HE]} \BibitemShut {NoStop}%
\bibitem [{\citenamefont {Zic}\ \emph {et~al.}(2023)\citenamefont {Zic} \emph {et~al.}}]{Zic:2023gta}%
  \BibitemOpen
  \bibfield  {author} {\bibinfo {author} {\bibfnamefont {Andrew}\ \bibnamefont {Zic}} \emph {et~al.},\ }\bibfield  {title} {\enquote {\bibinfo {title} {{The Parkes Pulsar Timing Array third data release}},}\ }\href {\doibase 10.1017/pasa.2023.36} {\bibfield  {journal} {\bibinfo  {journal} {Publ. Astron. Soc. Austral.}\ }\textbf {\bibinfo {volume} {40}},\ \bibinfo {pages} {e049} (\bibinfo {year} {2023})},\ \Eprint {http://arxiv.org/abs/2306.16230} {arXiv:2306.16230 [astro-ph.HE]} \BibitemShut {NoStop}%
\bibitem [{\citenamefont {Agazie}\ \emph {et~al.}(2023{\natexlab{b}})\citenamefont {Agazie}, \citenamefont {Anumarlapudi}, \citenamefont {Archibald} \emph {et~al.}}]{NANOGrav:2023hfp}%
  \BibitemOpen
  \bibfield  {author} {\bibinfo {author} {\bibfnamefont {Gabriella}\ \bibnamefont {Agazie}}, \bibinfo {author} {\bibfnamefont {Akash}\ \bibnamefont {Anumarlapudi}}, \bibinfo {author} {\bibfnamefont {Anne~M.}\ \bibnamefont {Archibald}},  \emph {et~al.} (\bibinfo {collaboration} {NANOGrav}),\ }\bibfield  {title} {\enquote {\bibinfo {title} {{The NANOGrav 15 yr Data Set: Constraints on Supermassive Black Hole Binaries from the Gravitational-wave Background}},}\ }\href {\doibase 10.3847/2041-8213/ace18b} {\  (\bibinfo {year} {2023}{\natexlab{b}}),\ 10.3847/2041-8213/ace18b},\ \Eprint {http://arxiv.org/abs/2306.16220} {arXiv:2306.16220 [astro-ph.HE]} \BibitemShut {NoStop}%
\bibitem [{\citenamefont {Sesana}\ \emph {et~al.}(2008)\citenamefont {Sesana}, \citenamefont {Vecchio},\ and\ \citenamefont {Colacino}}]{Sesana:2008mz}%
  \BibitemOpen
  \bibfield  {author} {\bibinfo {author} {\bibfnamefont {Alberto}\ \bibnamefont {Sesana}}, \bibinfo {author} {\bibfnamefont {Alberto}\ \bibnamefont {Vecchio}}, \ and\ \bibinfo {author} {\bibfnamefont {Carlo~Nicola}\ \bibnamefont {Colacino}},\ }\bibfield  {title} {\enquote {\bibinfo {title} {{The stochastic gravitational-wave background from massive black hole binary systems: implications for observations with Pulsar Timing Arrays}},}\ }\href {\doibase 10.1111/j.1365-2966.2008.13682.x} {\bibfield  {journal} {\bibinfo  {journal} {Mon. Not. Roy. Astron. Soc.}\ }\textbf {\bibinfo {volume} {390}},\ \bibinfo {pages} {192} (\bibinfo {year} {2008})},\ \Eprint {http://arxiv.org/abs/0804.4476} {arXiv:0804.4476 [astro-ph]} \BibitemShut {NoStop}%
\bibitem [{\citenamefont {Rosado}\ \emph {et~al.}(2015)\citenamefont {Rosado}, \citenamefont {Sesana},\ and\ \citenamefont {Gair}}]{Rosado:2015epa}%
  \BibitemOpen
  \bibfield  {author} {\bibinfo {author} {\bibfnamefont {Pablo~A.}\ \bibnamefont {Rosado}}, \bibinfo {author} {\bibfnamefont {Alberto}\ \bibnamefont {Sesana}}, \ and\ \bibinfo {author} {\bibfnamefont {Jonathan}\ \bibnamefont {Gair}},\ }\bibfield  {title} {\enquote {\bibinfo {title} {{Expected properties of the first gravitational wave signal detected with pulsar timing arrays}},}\ }\href {\doibase 10.1093/mnras/stv1098} {\bibfield  {journal} {\bibinfo  {journal} {Mon. Not. Roy. Astron. Soc.}\ }\textbf {\bibinfo {volume} {451}},\ \bibinfo {pages} {2417--2433} (\bibinfo {year} {2015})},\ \Eprint {http://arxiv.org/abs/1503.04803} {arXiv:1503.04803 [astro-ph.HE]} \BibitemShut {NoStop}%
\bibitem [{\citenamefont {Siemens}\ \emph {et~al.}(2019)\citenamefont {Siemens}, \citenamefont {Hazboun}, \citenamefont {Baker} \emph {et~al.}}]{Siemens:2019xkk}%
  \BibitemOpen
  \bibfield  {author} {\bibinfo {author} {\bibfnamefont {Xavier}\ \bibnamefont {Siemens}}, \bibinfo {author} {\bibfnamefont {Jeffrey~S.}\ \bibnamefont {Hazboun}}, \bibinfo {author} {\bibfnamefont {Paul~T.}\ \bibnamefont {Baker}},  \emph {et~al.},\ }\bibfield  {title} {\enquote {\bibinfo {title} {{Physics Beyond the Standard Model With Pulsar Timing Arrays}},}\ }\href@noop {} {\  (\bibinfo {year} {2019})},\ \Eprint {http://arxiv.org/abs/1907.04960} {arXiv:1907.04960 [gr-qc]} \BibitemShut {NoStop}%
\bibitem [{\citenamefont {Armstrong}\ \emph {et~al.}(2003)\citenamefont {Armstrong}, \citenamefont {Iess}, \citenamefont {Tortora},\ and\ \citenamefont {Bertotti}}]{Armstrong:2003ay}%
  \BibitemOpen
  \bibfield  {author} {\bibinfo {author} {\bibfnamefont {J.~W.}\ \bibnamefont {Armstrong}}, \bibinfo {author} {\bibfnamefont {L.}~\bibnamefont {Iess}}, \bibinfo {author} {\bibfnamefont {P.}~\bibnamefont {Tortora}}, \ and\ \bibinfo {author} {\bibfnamefont {B.}~\bibnamefont {Bertotti}},\ }\bibfield  {title} {\enquote {\bibinfo {title} {{Stochastic Gravitational Wave Background: Upper Limits in the $10^{-6}$ to $10^{-3}$ Hz Band}},}\ }\href {\doibase 10.1086/379505} {\bibfield  {journal} {\bibinfo  {journal} {Astrophys. J.}\ }\textbf {\bibinfo {volume} {599}},\ \bibinfo {pages} {806--813} (\bibinfo {year} {2003})}\BibitemShut {NoStop}%
\bibitem [{\citenamefont {Fedderke}\ \emph {et~al.}(2022)\citenamefont {Fedderke}, \citenamefont {Graham},\ and\ \citenamefont {Rajendran}}]{Fedderke:2021kuy}%
  \BibitemOpen
  \bibfield  {author} {\bibinfo {author} {\bibfnamefont {Michael~A.}\ \bibnamefont {Fedderke}}, \bibinfo {author} {\bibfnamefont {Peter~W.}\ \bibnamefont {Graham}}, \ and\ \bibinfo {author} {\bibfnamefont {Surjeet}\ \bibnamefont {Rajendran}},\ }\bibfield  {title} {\enquote {\bibinfo {title} {{Asteroids for $\mu$Hz gravitational-wave detection}},}\ }\href {\doibase 10.1103/PhysRevD.105.103018} {\bibfield  {journal} {\bibinfo  {journal} {Phys. Rev. D}\ }\textbf {\bibinfo {volume} {105}},\ \bibinfo {pages} {103018} (\bibinfo {year} {2022})},\ \Eprint {http://arxiv.org/abs/2112.11431} {arXiv:2112.11431 [gr-qc]} \BibitemShut {NoStop}%
\bibitem [{\citenamefont {Bustamante-Rosell}\ \emph {et~al.}(2022)\citenamefont {Bustamante-Rosell}, \citenamefont {Meyers}, \citenamefont {Pearson}, \citenamefont {Trendafilova},\ and\ \citenamefont {Zimmerman}}]{Bustamante-Rosell:2021daj}%
  \BibitemOpen
  \bibfield  {author} {\bibinfo {author} {\bibfnamefont {Mar{\'\i}a~Jos{\'e}}\ \bibnamefont {Bustamante-Rosell}}, \bibinfo {author} {\bibfnamefont {Joel}\ \bibnamefont {Meyers}}, \bibinfo {author} {\bibfnamefont {Noah}\ \bibnamefont {Pearson}}, \bibinfo {author} {\bibfnamefont {Cynthia}\ \bibnamefont {Trendafilova}}, \ and\ \bibinfo {author} {\bibfnamefont {Aaron}\ \bibnamefont {Zimmerman}},\ }\bibfield  {title} {\enquote {\bibinfo {title} {{Gravitational wave timing array}},}\ }\href {\doibase 10.1103/PhysRevD.105.044005} {\bibfield  {journal} {\bibinfo  {journal} {Phys. Rev. D}\ }\textbf {\bibinfo {volume} {105}},\ \bibinfo {pages} {044005} (\bibinfo {year} {2022})},\ \Eprint {http://arxiv.org/abs/2107.02788} {arXiv:2107.02788 [gr-qc]} \BibitemShut {NoStop}%
\bibitem [{\citenamefont {Sesana}\ \emph {et~al.}(2021)\citenamefont {Sesana} \emph {et~al.}}]{Sesana:2019vho}%
  \BibitemOpen
  \bibfield  {author} {\bibinfo {author} {\bibfnamefont {Alberto}\ \bibnamefont {Sesana}} \emph {et~al.},\ }\bibfield  {title} {\enquote {\bibinfo {title} {{Unveiling the gravitational universe at $\mu$-Hz frequencies}},}\ }\href {\doibase 10.1007/s10686-021-09709-9} {\bibfield  {journal} {\bibinfo  {journal} {Exper. Astron.}\ }\textbf {\bibinfo {volume} {51}},\ \bibinfo {pages} {1333--1383} (\bibinfo {year} {2021})},\ \Eprint {http://arxiv.org/abs/1908.11391} {arXiv:1908.11391 [astro-ph.IM]} \BibitemShut {NoStop}%
\bibitem [{\citenamefont {Begelman}\ \emph {et~al.}(1980)\citenamefont {Begelman}, \citenamefont {Blandford},\ and\ \citenamefont {Rees}}]{Begelman:1980vb}%
  \BibitemOpen
  \bibfield  {author} {\bibinfo {author} {\bibfnamefont {M.~C.}\ \bibnamefont {Begelman}}, \bibinfo {author} {\bibfnamefont {R.~D.}\ \bibnamefont {Blandford}}, \ and\ \bibinfo {author} {\bibfnamefont {M.~J.}\ \bibnamefont {Rees}},\ }\bibfield  {title} {\enquote {\bibinfo {title} {{Massive black hole binaries in active galactic nuclei}},}\ }\href {\doibase 10.1038/287307a0} {\bibfield  {journal} {\bibinfo  {journal} {Nature}\ }\textbf {\bibinfo {volume} {287}},\ \bibinfo {pages} {307--309} (\bibinfo {year} {1980})}\BibitemShut {NoStop}%
\bibitem [{\citenamefont {Phinney}(2001)}]{Phinney:2001di}%
  \BibitemOpen
  \bibfield  {author} {\bibinfo {author} {\bibfnamefont {E.~S.}\ \bibnamefont {Phinney}},\ }\bibfield  {title} {\enquote {\bibinfo {title} {{A Practical theorem on gravitational wave backgrounds}},}\ }\href@noop {} {\  (\bibinfo {year} {2001})},\ \Eprint {http://arxiv.org/abs/astro-ph/0108028} {arXiv:astro-ph/0108028} \BibitemShut {NoStop}%
\bibitem [{\citenamefont {Burke-Spolaor}\ \emph {et~al.}(2019)\citenamefont {Burke-Spolaor} \emph {et~al.}}]{Burke-Spolaor:2018bvk}%
  \BibitemOpen
  \bibfield  {author} {\bibinfo {author} {\bibfnamefont {Sarah}\ \bibnamefont {Burke-Spolaor}} \emph {et~al.},\ }\bibfield  {title} {\enquote {\bibinfo {title} {{The Astrophysics of Nanohertz Gravitational Waves}},}\ }\href {\doibase 10.1007/s00159-019-0115-7} {\bibfield  {journal} {\bibinfo  {journal} {Astron. Astrophys. Rev.}\ }\textbf {\bibinfo {volume} {27}},\ \bibinfo {pages} {5} (\bibinfo {year} {2019})},\ \Eprint {http://arxiv.org/abs/1811.08826} {arXiv:1811.08826 [astro-ph.HE]} \BibitemShut {NoStop}%
\bibitem [{\citenamefont {Afzal}\ \emph {et~al.}(2023)\citenamefont {Afzal}, \citenamefont {Agazie}, \citenamefont {Anumarlapudi} \emph {et~al.}}]{NANOGrav:2023hvm}%
  \BibitemOpen
  \bibfield  {author} {\bibinfo {author} {\bibfnamefont {Adeela}\ \bibnamefont {Afzal}}, \bibinfo {author} {\bibfnamefont {Gabriella}\ \bibnamefont {Agazie}}, \bibinfo {author} {\bibfnamefont {Akash}\ \bibnamefont {Anumarlapudi}},  \emph {et~al.} (\bibinfo {collaboration} {NANOGrav}),\ }\bibfield  {title} {\enquote {\bibinfo {title} {{The NANOGrav 15 yr Data Set: Search for Signals from New Physics}},}\ }\href {\doibase 10.3847/2041-8213/acdc91} {\bibfield  {journal} {\bibinfo  {journal} {Astrophys. J. Lett.}\ }\textbf {\bibinfo {volume} {951}},\ \bibinfo {pages} {L11} (\bibinfo {year} {2023})},\ \Eprint {http://arxiv.org/abs/2306.16219} {arXiv:2306.16219 [astro-ph.HE]} \BibitemShut {NoStop}%
\bibitem [{\citenamefont {Book}\ and\ \citenamefont {Flanagan}()}]{Book:2010pf}%
  \BibitemOpen
  \bibfield  {author} {\bibinfo {author} {\bibfnamefont {Laura~G.}\ \bibnamefont {Book}}\ and\ \bibinfo {author} {\bibfnamefont {Eanna~E.}\ \bibnamefont {Flanagan}},\ }\bibfield  {title} {\enquote {\bibinfo {title} {{Astrometric Effects of a Stochastic Gravitational Wave Background}},}\ }\href {\doibase 10.1103/PhysRevD.83.024024} {\ 10.1103/PhysRevD.83.024024},\ \Eprint {http://arxiv.org/abs/1009.4192} {arXiv:1009.4192 [astro-ph.CO]} \BibitemShut {NoStop}%
\bibitem [{\citenamefont {Darling}\ \emph {et~al.}(2018)\citenamefont {Darling}, \citenamefont {Truebenbach},\ and\ \citenamefont {Paine}}]{Darling:2018hmc}%
  \BibitemOpen
  \bibfield  {author} {\bibinfo {author} {\bibfnamefont {Jeremy}\ \bibnamefont {Darling}}, \bibinfo {author} {\bibfnamefont {Alexandra~E.}\ \bibnamefont {Truebenbach}}, \ and\ \bibinfo {author} {\bibfnamefont {Jennie}\ \bibnamefont {Paine}},\ }\bibfield  {title} {\enquote {\bibinfo {title} {{Astrometric Limits on the Stochastic Gravitational Wave Background}},}\ }\href {\doibase 10.3847/1538-4357/aac772} {\  (\bibinfo {year} {2018}),\ 10.3847/1538-4357/aac772},\ \Eprint {http://arxiv.org/abs/1804.06986} {arXiv:1804.06986 [astro-ph.IM]} \BibitemShut {NoStop}%
\bibitem [{\citenamefont {Wang}\ \emph {et~al.}(2022)\citenamefont {Wang}, \citenamefont {Pardo}, \citenamefont {Chang},\ and\ \citenamefont {Doré}}]{Wang:2022sxn}%
  \BibitemOpen
  \bibfield  {author} {\bibinfo {author} {\bibfnamefont {Yijun}\ \bibnamefont {Wang}}, \bibinfo {author} {\bibfnamefont {Kris}\ \bibnamefont {Pardo}}, \bibinfo {author} {\bibfnamefont {Tzu-Ching}\ \bibnamefont {Chang}}, \ and\ \bibinfo {author} {\bibfnamefont {Olivier}\ \bibnamefont {Doré}},\ }\bibfield  {title} {\enquote {\bibinfo {title} {{Constraining the stochastic gravitational wave background with photometric surveys}},}\ }\href {\doibase 10.1103/PhysRevD.106.084006} {\bibfield  {journal} {\bibinfo  {journal} {Phys. Rev. D}\ }\textbf {\bibinfo {volume} {106}},\ \bibinfo {pages} {084006} (\bibinfo {year} {2022})},\ \Eprint {http://arxiv.org/abs/2205.07962} {arXiv:2205.07962 [gr-qc]} \BibitemShut {NoStop}%
\bibitem [{\citenamefont {Pardo}\ \emph {et~al.}(2023)\citenamefont {Pardo}, \citenamefont {Chang}, \citenamefont {Doré},\ and\ \citenamefont {Wang}}]{Pardo:2023cag}%
  \BibitemOpen
  \bibfield  {author} {\bibinfo {author} {\bibfnamefont {Kris}\ \bibnamefont {Pardo}}, \bibinfo {author} {\bibfnamefont {Tzu-Ching}\ \bibnamefont {Chang}}, \bibinfo {author} {\bibfnamefont {Olivier}\ \bibnamefont {Doré}}, \ and\ \bibinfo {author} {\bibfnamefont {Yijun}\ \bibnamefont {Wang}},\ }\bibfield  {title} {\enquote {\bibinfo {title} {{Gravitational Wave Detection with Relative Astrometry using Roman's Galactic Bulge Time Domain Survey}},}\ }\href@noop {} {\  (\bibinfo {year} {2023})},\ \Eprint {http://arxiv.org/abs/2306.14968} {arXiv:2306.14968 [astro-ph.GA]} \BibitemShut {NoStop}%
\bibitem [{\citenamefont {Zhang}\ \emph {et~al.}(2025)\citenamefont {Zhang}, \citenamefont {Pardo}, \citenamefont {Wang}, \citenamefont {Bouma}, \citenamefont {Chang},\ and\ \citenamefont {Dor{\'e}}}]{Zhang:2025srs}%
  \BibitemOpen
  \bibfield  {author} {\bibinfo {author} {\bibfnamefont {Benjamin}\ \bibnamefont {Zhang}}, \bibinfo {author} {\bibfnamefont {Kris}\ \bibnamefont {Pardo}}, \bibinfo {author} {\bibfnamefont {Yijun}\ \bibnamefont {Wang}}, \bibinfo {author} {\bibfnamefont {Luke}\ \bibnamefont {Bouma}}, \bibinfo {author} {\bibfnamefont {Tzu-Ching}\ \bibnamefont {Chang}}, \ and\ \bibinfo {author} {\bibfnamefont {Olivier}\ \bibnamefont {Dor{\'e}}},\ }\bibfield  {title} {\enquote {\bibinfo {title} {{Fast Bayesian method for coherent gravitational wave searches with relative astrometry}},}\ }\href {\doibase 10.1103/qzys-t232} {\bibfield  {journal} {\bibinfo  {journal} {Phys. Rev. D}\ }\textbf {\bibinfo {volume} {112}},\ \bibinfo {pages} {042002} (\bibinfo {year} {2025})},\ \Eprint {http://arxiv.org/abs/2506.19206} {arXiv:2506.19206 [astro-ph.IM]} \BibitemShut {NoStop}%
\bibitem [{\citenamefont {Schmidt}\ and\ \citenamefont {Jeong}({\natexlab{a}})}]{Schmidt:2012nw}%
  \BibitemOpen
  \bibfield  {author} {\bibinfo {author} {\bibfnamefont {Fabian}\ \bibnamefont {Schmidt}}\ and\ \bibinfo {author} {\bibfnamefont {Donghui}\ \bibnamefont {Jeong}},\ }\bibfield  {title} {\enquote {\bibinfo {title} {{Large-Scale Structure with Gravitational Waves II: Shear}},}\ }\href {\doibase 10.1103/PhysRevD.86.083513} {\  ({\natexlab{a}}),\ 10.1103/PhysRevD.86.083513},\ \Eprint {http://arxiv.org/abs/1205.1514} {arXiv:1205.1514 [astro-ph.CO]} \BibitemShut {NoStop}%
\bibitem [{\citenamefont {Schmidt}\ and\ \citenamefont {Jeong}({\natexlab{b}})}]{Schmidt:2012ne}%
  \BibitemOpen
  \bibfield  {author} {\bibinfo {author} {\bibfnamefont {Fabian}\ \bibnamefont {Schmidt}}\ and\ \bibinfo {author} {\bibfnamefont {Donghui}\ \bibnamefont {Jeong}},\ }\bibfield  {title} {\enquote {\bibinfo {title} {{Cosmic Rulers}},}\ }\href {\doibase 10.1103/PhysRevD.86.083527} {\  ({\natexlab{b}}),\ 10.1103/PhysRevD.86.083527},\ \Eprint {http://arxiv.org/abs/1204.3625} {arXiv:1204.3625 [astro-ph.CO]} \BibitemShut {NoStop}%
\bibitem [{\citenamefont {Jeong}\ and\ \citenamefont {Schmidt}()}]{Jeong:2012nu}%
  \BibitemOpen
  \bibfield  {author} {\bibinfo {author} {\bibfnamefont {Donghui}\ \bibnamefont {Jeong}}\ and\ \bibinfo {author} {\bibfnamefont {Fabian}\ \bibnamefont {Schmidt}},\ }\bibfield  {title} {\enquote {\bibinfo {title} {{Large-Scale Structure with Gravitational Waves I: Galaxy Clustering}},}\ }\href {\doibase 10.1103/PhysRevD.86.083512} {\ 10.1103/PhysRevD.86.083512},\ \Eprint {http://arxiv.org/abs/1205.1512} {arXiv:1205.1512 [astro-ph.CO]} \BibitemShut {NoStop}%
\bibitem [{\citenamefont {Schmidt}\ \emph {et~al.}(2013)\citenamefont {Schmidt}, \citenamefont {Pajer},\ and\ \citenamefont {Zaldarriaga}}]{Schmidt:2013gwa}%
  \BibitemOpen
  \bibfield  {author} {\bibinfo {author} {\bibfnamefont {Fabian}\ \bibnamefont {Schmidt}}, \bibinfo {author} {\bibfnamefont {Enrico}\ \bibnamefont {Pajer}}, \ and\ \bibinfo {author} {\bibfnamefont {Matias}\ \bibnamefont {Zaldarriaga}},\ }\bibfield  {title} {\enquote {\bibinfo {title} {{Large-Scale Structure and Gravitational Waves III: Tidal Effects}},}\ }\href {\doibase 10.1103/PhysRevD.89.083507} {\  (\bibinfo {year} {2013}),\ 10.1103/PhysRevD.89.083507},\ \Eprint {http://arxiv.org/abs/1312.5616} {arXiv:1312.5616 [astro-ph.CO]} \BibitemShut {NoStop}%
\bibitem [{\citenamefont {Mentasti}\ and\ \citenamefont {Contaldi}(2024)}]{Mentasti:2024fgt}%
  \BibitemOpen
  \bibfield  {author} {\bibinfo {author} {\bibfnamefont {Giorgio}\ \bibnamefont {Mentasti}}\ and\ \bibinfo {author} {\bibfnamefont {Carlo~R.}\ \bibnamefont {Contaldi}},\ }\bibfield  {title} {\enquote {\bibinfo {title} {{Cosmic shimmering: the gravitational wave signal of time-resolved cosmic shear observations}},}\ }\href {\doibase 10.1088/1475-7516/2025/06/013} {\  (\bibinfo {year} {2024}),\ 10.1088/1475-7516/2025/06/013},\ \Eprint {http://arxiv.org/abs/2410.16274} {arXiv:2410.16274 [astro-ph.CO]} \BibitemShut {NoStop}%
\bibitem [{\citenamefont {Troxel}\ \emph {et~al.}(2022)\citenamefont {Troxel}, \citenamefont {Lin}, \citenamefont {Park} \emph {et~al.}}]{LSSTDarkEnergyScience:2022nlw}%
  \BibitemOpen
  \bibfield  {author} {\bibinfo {author} {\bibfnamefont {M.A.}\ \bibnamefont {Troxel}}, \bibinfo {author} {\bibfnamefont {C.}~\bibnamefont {Lin}}, \bibinfo {author} {\bibfnamefont {A.}~\bibnamefont {Park}},  \emph {et~al.} (\bibinfo {collaboration} {LSST Dark Energy Science}),\ }\bibfield  {title} {\enquote {\bibinfo {title} {{A joint Roman Space Telescope and Rubin Observatory synthetic wide-field imaging survey}},}\ }\href {\doibase 10.1093/mnras/stad664} {\  (\bibinfo {year} {2022}),\ 10.1093/mnras/stad664},\ \Eprint {http://arxiv.org/abs/2209.06829} {arXiv:2209.06829 [astro-ph.IM]} \BibitemShut {NoStop}%
\bibitem [{\citenamefont {Spergel}\ \emph {et~al.}(2015)\citenamefont {Spergel} \emph {et~al.}}]{Spergel:2015sza}%
  \BibitemOpen
  \bibfield  {author} {\bibinfo {author} {\bibfnamefont {D.}~\bibnamefont {Spergel}} \emph {et~al.},\ }\bibfield  {title} {\enquote {\bibinfo {title} {{Wide-Field InfrarRed Survey Telescope-Astrophysics Focused Telescope Assets WFIRST-AFTA 2015 Report}},}\ }\href@noop {} {\  (\bibinfo {year} {2015})},\ \Eprint {http://arxiv.org/abs/1503.03757} {arXiv:1503.03757 [astro-ph.IM]} \BibitemShut {NoStop}%
\bibitem [{\citenamefont {Laureijs}\ \emph {et~al.}(2011)\citenamefont {Laureijs} \emph {et~al.}}]{EUCLID:2011zbd}%
  \BibitemOpen
  \bibfield  {author} {\bibinfo {author} {\bibfnamefont {R.}~\bibnamefont {Laureijs}} \emph {et~al.} (\bibinfo {collaboration} {EUCLID}),\ }\bibfield  {title} {\enquote {\bibinfo {title} {{Euclid Definition Study Report}},}\ }\href@noop {} {\  (\bibinfo {year} {2011})},\ \Eprint {http://arxiv.org/abs/1110.3193} {arXiv:1110.3193 [astro-ph.CO]} \BibitemShut {NoStop}%
\bibitem [{\citenamefont {Mellier}\ \emph {et~al.}(2024)\citenamefont {Mellier}, \citenamefont {Barroso}, \citenamefont {Achúcarro} \emph {et~al.}}]{Euclid:2024yrr}%
  \BibitemOpen
  \bibfield  {author} {\bibinfo {author} {\bibfnamefont {Y.}~\bibnamefont {Mellier}}, \bibinfo {author} {\bibfnamefont {J.A.~Acevedo}\ \bibnamefont {Barroso}}, \bibinfo {author} {\bibfnamefont {A.}~\bibnamefont {Achúcarro}},  \emph {et~al.} (\bibinfo {collaboration} {Euclid}),\ }\bibfield  {title} {\enquote {\bibinfo {title} {{Euclid. I. Overview of the Euclid mission}},}\ }\href {\doibase 10.1051/0004-6361/202450810} {\  (\bibinfo {year} {2024}),\ 10.1051/0004-6361/202450810},\ \Eprint {http://arxiv.org/abs/2405.13491} {arXiv:2405.13491 [astro-ph.CO]} \BibitemShut {NoStop}%
\bibitem [{\citenamefont {Zeghal}\ \emph {et~al.}(2025)\citenamefont {Zeghal}, \citenamefont {Lanzieri}, \citenamefont {Lanusse} \emph {et~al.}}]{Zeghal:2024kic}%
  \BibitemOpen
  \bibfield  {author} {\bibinfo {author} {\bibfnamefont {Justine}\ \bibnamefont {Zeghal}}, \bibinfo {author} {\bibfnamefont {Denise}\ \bibnamefont {Lanzieri}}, \bibinfo {author} {\bibfnamefont {François}\ \bibnamefont {Lanusse}},  \emph {et~al.} (\bibinfo {collaboration} {LSST Dark Energy Science}),\ }\bibfield  {title} {\enquote {\bibinfo {title} {{Simulation-based inference benchmark for weak lensing cosmology}},}\ }\href {\doibase 10.1051/0004-6361/202452410} {\bibfield  {journal} {\bibinfo  {journal} {Astron. Astrophys.}\ }\textbf {\bibinfo {volume} {699}},\ \bibinfo {pages} {A327} (\bibinfo {year} {2025})},\ \Eprint {http://arxiv.org/abs/2409.17975} {arXiv:2409.17975 [astro-ph.CO]} \BibitemShut {NoStop}%
\bibitem [{\citenamefont {Bera}\ \emph {et~al.}(2025)\citenamefont {Bera}, \citenamefont {Varela}, \citenamefont {Sooriyaarachchi} \emph {et~al.}}]{Bera:2025ixc}%
  \BibitemOpen
  \bibfield  {author} {\bibinfo {author} {\bibfnamefont {Avijit}\ \bibnamefont {Bera}}, \bibinfo {author} {\bibfnamefont {Leonel~Medina}\ \bibnamefont {Varela}}, \bibinfo {author} {\bibfnamefont {Vinu}\ \bibnamefont {Sooriyaarachchi}},  \emph {et~al.} (\bibinfo {collaboration} {LSST Dark Energy Science}),\ }\bibfield  {title} {\enquote {\bibinfo {title} {{A direct detection method of galaxy intrinsic ellipticity-gravitational shear correlation in non-linear regimes using self-calibration}},}\ }\href {\doibase 10.1088/1475-7516/2025/09/038} {\  (\bibinfo {year} {2025}),\ 10.1088/1475-7516/2025/09/038},\ \Eprint {http://arxiv.org/abs/2503.24269} {arXiv:2503.24269 [astro-ph.CO]} \BibitemShut {NoStop}%
\bibitem [{\citenamefont {Dodelson}\ and\ \citenamefont {Schmidt}(2020)}]{Dodelson:2020bqr}%
  \BibitemOpen
  \bibfield  {author} {\bibinfo {author} {\bibfnamefont {Scott}\ \bibnamefont {Dodelson}}\ and\ \bibinfo {author} {\bibfnamefont {Fabian}\ \bibnamefont {Schmidt}},\ }\href {\doibase 10.1016/C2017-0-01943-2} {\emph {\bibinfo {title} {{Modern Cosmology}}}}\ (\bibinfo  {publisher} {Academic Press},\ \bibinfo {year} {2020})\BibitemShut {NoStop}%
\bibitem [{\citenamefont {Jackson}(1998)}]{Jackson:1998nia}%
  \BibitemOpen
  \bibfield  {author} {\bibinfo {author} {\bibfnamefont {John~David}\ \bibnamefont {Jackson}},\ }\href@noop {} {\emph {\bibinfo {title} {{Classical Electrodynamics}}}}\ (\bibinfo  {publisher} {Wiley},\ \bibinfo {year} {1998})\BibitemShut {NoStop}%
\bibitem [{\citenamefont {Mihaylov}\ \emph {et~al.}(2018)\citenamefont {Mihaylov}, \citenamefont {Moore}, \citenamefont {Gair} \emph {et~al.}}]{Mihaylov:2018uqm}%
  \BibitemOpen
  \bibfield  {author} {\bibinfo {author} {\bibfnamefont {Deyan~P.}\ \bibnamefont {Mihaylov}}, \bibinfo {author} {\bibfnamefont {Christopher~J.}\ \bibnamefont {Moore}}, \bibinfo {author} {\bibfnamefont {Jonathan~R.}\ \bibnamefont {Gair}},  \emph {et~al.},\ }\bibfield  {title} {\enquote {\bibinfo {title} {{Astrometric Effects of Gravitational Wave Backgrounds with non-Einsteinian Polarizations}},}\ }\href {\doibase 10.1103/PhysRevD.97.124058} {\  (\bibinfo {year} {2018}),\ 10.1103/PhysRevD.97.124058},\ \Eprint {http://arxiv.org/abs/1804.00660} {arXiv:1804.00660 [gr-qc]} \BibitemShut {NoStop}%
\bibitem [{\citenamefont {Madison}(2020)}]{Madison:2020xhh}%
  \BibitemOpen
  \bibfield  {author} {\bibinfo {author} {\bibfnamefont {Dustin~R.}\ \bibnamefont {Madison}},\ }\bibfield  {title} {\enquote {\bibinfo {title} {{Persistent Astrometric Deflections from Gravitational-Wave Memory}},}\ }\href {\doibase 10.1103/PhysRevLett.125.041101} {\bibfield  {journal} {\bibinfo  {journal} {Phys. Rev. Lett.}\ }\textbf {\bibinfo {volume} {125}},\ \bibinfo {pages} {041101} (\bibinfo {year} {2020})},\ \Eprint {http://arxiv.org/abs/2007.12206} {arXiv:2007.12206 [gr-qc]} \BibitemShut {NoStop}%
\bibitem [{\citenamefont {Chang}\ \emph {et~al.}(2013)\citenamefont {Chang}, \citenamefont {Jarvis}, \citenamefont {Jain} \emph {et~al.}}]{Chang:2013xja}%
  \BibitemOpen
  \bibfield  {author} {\bibinfo {author} {\bibfnamefont {C.}~\bibnamefont {Chang}}, \bibinfo {author} {\bibfnamefont {M.}~\bibnamefont {Jarvis}}, \bibinfo {author} {\bibfnamefont {B.}~\bibnamefont {Jain}},  \emph {et~al.},\ }\bibfield  {title} {\enquote {\bibinfo {title} {{The Effective Number Density of Galaxies for Weak Lensing Measurements in the LSST Project}},}\ }\href {\doibase 10.1093/mnras/stt1156} {\  (\bibinfo {year} {2013}),\ 10.1093/mnras/stt1156},\ \Eprint {http://arxiv.org/abs/1305.0793} {arXiv:1305.0793 [astro-ph.CO]} \BibitemShut {NoStop}%
\bibitem [{\citenamefont {Albrecht}\ \emph {et~al.}()\citenamefont {Albrecht}, \citenamefont {Bernstein}, \citenamefont {Cahn} \emph {et~al.}}]{Albrecht:2006um}%
  \BibitemOpen
  \bibfield  {author} {\bibinfo {author} {\bibfnamefont {Andreas}\ \bibnamefont {Albrecht}}, \bibinfo {author} {\bibfnamefont {Gary}\ \bibnamefont {Bernstein}}, \bibinfo {author} {\bibfnamefont {Robert}\ \bibnamefont {Cahn}},  \emph {et~al.},\ }\bibfield  {title} {\enquote {\bibinfo {title} {{Report of the Dark Energy Task Force}},}\ }\href@noop {} {\ }\Eprint {http://arxiv.org/abs/astro-ph/0609591} {arXiv:astro-ph/0609591 [astro-ph]} \BibitemShut {NoStop}%
\bibitem [{\citenamefont {Moore}\ \emph {et~al.}(2014)\citenamefont {Moore}, \citenamefont {Cole},\ and\ \citenamefont {Berry}}]{Moore:2014lga}%
  \BibitemOpen
  \bibfield  {author} {\bibinfo {author} {\bibfnamefont {C.J.}\ \bibnamefont {Moore}}, \bibinfo {author} {\bibfnamefont {R.H.}\ \bibnamefont {Cole}}, \ and\ \bibinfo {author} {\bibfnamefont {C.P.L.}\ \bibnamefont {Berry}},\ }\bibfield  {title} {\enquote {\bibinfo {title} {{Gravitational-wave sensitivity curves}},}\ }\href {\doibase 10.1088/0264-9381/32/1/015014} {\  (\bibinfo {year} {2014}),\ 10.1088/0264-9381/32/1/015014},\ \Eprint {http://arxiv.org/abs/1408.0740} {arXiv:1408.0740 [gr-qc]} \BibitemShut {NoStop}%
\bibitem [{\citenamefont {Hogg}()}]{Hogg:1999ad}%
  \BibitemOpen
  \bibfield  {author} {\bibinfo {author} {\bibfnamefont {David~W.}\ \bibnamefont {Hogg}},\ }\bibfield  {title} {\enquote {\bibinfo {title} {{Distance measures in cosmology}},}\ }\href@noop {} {\ }\Eprint {http://arxiv.org/abs/astro-ph/9905116} {arXiv:astro-ph/9905116 [astro-ph]} \BibitemShut {NoStop}%
\bibitem [{\citenamefont {Schechter}(1976)}]{Schechter:1976iz}%
  \BibitemOpen
  \bibfield  {author} {\bibinfo {author} {\bibfnamefont {P.}~\bibnamefont {Schechter}},\ }\bibfield  {title} {\enquote {\bibinfo {title} {{An analytic expression for the luminosity function for galaxies}},}\ }\href {\doibase 10.1086/154079} {\bibfield  {journal} {\bibinfo  {journal} {Astrophys. J.}\ }\textbf {\bibinfo {volume} {203}},\ \bibinfo {pages} {297--306} (\bibinfo {year} {1976})}\BibitemShut {NoStop}%
\bibitem [{\citenamefont {Loveday}\ \emph {et~al.}()\citenamefont {Loveday}, \citenamefont {Norberg}, \citenamefont {Baldry} \emph {et~al.}}]{Loveday:2011dh}%
  \BibitemOpen
  \bibfield  {author} {\bibinfo {author} {\bibfnamefont {J.}~\bibnamefont {Loveday}}, \bibinfo {author} {\bibfnamefont {P.}~\bibnamefont {Norberg}}, \bibinfo {author} {\bibfnamefont {I.K.}\ \bibnamefont {Baldry}},  \emph {et~al.},\ }\bibfield  {title} {\enquote {\bibinfo {title} {{Galaxy and Mass Assembly (GAMA): ugriz galaxy luminosity functions}},}\ }\href {\doibase 10.1111/j.1365-2966.2011.20111.x} {\ 10.1111/j.1365-2966.2011.20111.x},\ \Eprint {http://arxiv.org/abs/1111.0166} {arXiv:1111.0166 [astro-ph.CO]} \BibitemShut {NoStop}%
\bibitem [{\citenamefont {Ivezić}\ \emph {et~al.}()\citenamefont {Ivezić}, \citenamefont {Kahn}, \citenamefont {Tyson} \emph {et~al.}}]{LSST:2008ijt}%
  \BibitemOpen
  \bibfield  {author} {\bibinfo {author} {\bibfnamefont {Željko}\ \bibnamefont {Ivezić}}, \bibinfo {author} {\bibfnamefont {Steven~M.}\ \bibnamefont {Kahn}}, \bibinfo {author} {\bibfnamefont {J.Anthony}\ \bibnamefont {Tyson}},  \emph {et~al.} (\bibinfo {collaboration} {LSST}),\ }\bibfield  {title} {\enquote {\bibinfo {title} {{LSST: from Science Drivers to Reference Design and Anticipated Data Products}},}\ }\href {\doibase 10.3847/1538-4357/ab042c} {\ 10.3847/1538-4357/ab042c},\ \Eprint {http://arxiv.org/abs/0805.2366} {arXiv:0805.2366 [astro-ph]} \BibitemShut {NoStop}%
\bibitem [{\citenamefont {Shen}\ \emph {et~al.}(2003)\citenamefont {Shen}, \citenamefont {Mo}, \citenamefont {White}, \citenamefont {Blanton}, \citenamefont {Kauffmann}, \citenamefont {Voges}, \citenamefont {Brinkmann},\ and\ \citenamefont {Csabai}}]{Shen:2003sda}%
  \BibitemOpen
  \bibfield  {author} {\bibinfo {author} {\bibfnamefont {Shiyin}\ \bibnamefont {Shen}}, \bibinfo {author} {\bibfnamefont {H.~J.}\ \bibnamefont {Mo}}, \bibinfo {author} {\bibfnamefont {Simon D.~M.}\ \bibnamefont {White}}, \bibinfo {author} {\bibfnamefont {Michael~R.}\ \bibnamefont {Blanton}}, \bibinfo {author} {\bibfnamefont {Guinevere}\ \bibnamefont {Kauffmann}}, \bibinfo {author} {\bibfnamefont {Wolfgang}\ \bibnamefont {Voges}}, \bibinfo {author} {\bibfnamefont {J.}~\bibnamefont {Brinkmann}}, \ and\ \bibinfo {author} {\bibfnamefont {Istvan}\ \bibnamefont {Csabai}},\ }\bibfield  {title} {\enquote {\bibinfo {title} {{The Size distribution of galaxies in the Sloan Digital Sky Survey}},}\ }\href {\doibase 10.1046/j.1365-8711.2003.06740.x} {\bibfield  {journal} {\bibinfo  {journal} {Mon. Not. Roy. Astron. Soc.}\ }\textbf {\bibinfo {volume} {343}},\ \bibinfo {pages} {978} (\bibinfo {year} {2003})},\ \Eprint {http://arxiv.org/abs/astro-ph/0301527} {arXiv:astro-ph/0301527} \BibitemShut {NoStop}%
\bibitem [{\citenamefont {Cameron}\ and\ \citenamefont {Driver}(2007)}]{Cameron:2007sx}%
  \BibitemOpen
  \bibfield  {author} {\bibinfo {author} {\bibfnamefont {Ewan}\ \bibnamefont {Cameron}}\ and\ \bibinfo {author} {\bibfnamefont {S.~P.}\ \bibnamefont {Driver}},\ }\bibfield  {title} {\enquote {\bibinfo {title} {{The galaxy luminosity-size relation and selection biases in the Hubble Ultra Deep Field}},}\ }\href {\doibase 10.1111/j.1365-2966.2007.11507.x} {\bibfield  {journal} {\bibinfo  {journal} {Mon. Not. Roy. Astron. Soc.}\ }\textbf {\bibinfo {volume} {377}},\ \bibinfo {pages} {523--534} (\bibinfo {year} {2007})},\ \Eprint {http://arxiv.org/abs/astro-ph/0701419} {arXiv:astro-ph/0701419} \BibitemShut {NoStop}%
\bibitem [{\citenamefont {Newton}\ \emph {et~al.}()\citenamefont {Newton}, \citenamefont {Marshall}, \citenamefont {Auger} \emph {et~al.}}]{Newton:2011jw}%
  \BibitemOpen
  \bibfield  {author} {\bibinfo {author} {\bibfnamefont {Elisabeth~R.}\ \bibnamefont {Newton}}, \bibinfo {author} {\bibfnamefont {Philip~J.}\ \bibnamefont {Marshall}}, \bibinfo {author} {\bibfnamefont {Matthew~W.}\ \bibnamefont {Auger}},  \emph {et~al.},\ }\bibfield  {title} {\enquote {\bibinfo {title} {{The Sloan Lens ACS Survey. XI. Beyond Hubble resolution: size, luminosity and stellar mass of compact lensed galaxies at intermediate redshift}},}\ }\href {\doibase 10.1088/0004-637X/734/2/104} {\ 10.1088/0004-637X/734/2/104},\ \Eprint {http://arxiv.org/abs/1104.2608} {arXiv:1104.2608 [astro-ph.CO]} \BibitemShut {NoStop}%
\bibitem [{\citenamefont {Antoniadis}\ \emph {et~al.}(2023{\natexlab{b}})\citenamefont {Antoniadis}, \citenamefont {Arumugam}, \citenamefont {Arumugam} \emph {et~al.}}]{EPTA:2023xxk}%
  \BibitemOpen
  \bibfield  {author} {\bibinfo {author} {\bibfnamefont {J.}~\bibnamefont {Antoniadis}}, \bibinfo {author} {\bibfnamefont {P.}~\bibnamefont {Arumugam}}, \bibinfo {author} {\bibfnamefont {S.}~\bibnamefont {Arumugam}},  \emph {et~al.} (\bibinfo {collaboration} {EPTA}),\ }\bibfield  {title} {\enquote {\bibinfo {title} {{The second data release from the European Pulsar Timing Array - IV. Implications for massive black holes, dark matter, and the early Universe}},}\ }\href {\doibase 10.1051/0004-6361/202347433} {\  (\bibinfo {year} {2023}{\natexlab{b}}),\ 10.1051/0004-6361/202347433},\ \Eprint {http://arxiv.org/abs/2306.16227} {arXiv:2306.16227 [astro-ph.CO]} \BibitemShut {NoStop}%
\bibitem [{\citenamefont {Agazie}\ \emph {et~al.}(2023{\natexlab{c}})\citenamefont {Agazie}, \citenamefont {Anumarlapudi}, \citenamefont {Archibald} \emph {et~al.}}]{NANOGrav:2023pdq}%
  \BibitemOpen
  \bibfield  {author} {\bibinfo {author} {\bibfnamefont {Gabriella}\ \bibnamefont {Agazie}}, \bibinfo {author} {\bibfnamefont {Akash}\ \bibnamefont {Anumarlapudi}}, \bibinfo {author} {\bibfnamefont {Anne~M.}\ \bibnamefont {Archibald}},  \emph {et~al.} (\bibinfo {collaboration} {NANOGrav}),\ }\bibfield  {title} {\enquote {\bibinfo {title} {{The NANOGrav 15 yr Data Set: Bayesian Limits on Gravitational Waves from Individual Supermassive Black Hole Binaries}},}\ }\href {\doibase 10.3847/2041-8213/ace18a} {\  (\bibinfo {year} {2023}{\natexlab{c}}),\ 10.3847/2041-8213/ace18a},\ \Eprint {http://arxiv.org/abs/2306.16222} {arXiv:2306.16222 [astro-ph.HE]} \BibitemShut {NoStop}%
\bibitem [{\citenamefont {Tinto}\ and\ \citenamefont {Armstrong}(1998)}]{Tinto:1998ee}%
  \BibitemOpen
  \bibfield  {author} {\bibinfo {author} {\bibfnamefont {Massimo}\ \bibnamefont {Tinto}}\ and\ \bibinfo {author} {\bibfnamefont {J.~W.}\ \bibnamefont {Armstrong}},\ }\bibfield  {title} {\enquote {\bibinfo {title} {{Spacecraft Doppler Tracking as a Narrow Band Detector of Gravitational Radiation}},}\ }\href {\doibase 10.1103/PhysRevD.58.042002} {\bibfield  {journal} {\bibinfo  {journal} {Phys. Rev. D}\ }\textbf {\bibinfo {volume} {58}},\ \bibinfo {pages} {042002} (\bibinfo {year} {1998})}\BibitemShut {NoStop}%
\bibitem [{\citenamefont {Abbate}\ \emph {et~al.}(2003)\citenamefont {Abbate} \emph {et~al.}}]{Abbate:2003stu}%
  \BibitemOpen
  \bibfield  {author} {\bibinfo {author} {\bibfnamefont {Salvatore~F.}\ \bibnamefont {Abbate}} \emph {et~al.},\ }\bibfield  {title} {\enquote {\bibinfo {title} {{The Cassini Gravitational Wave Experiment}},}\ }\href {\doibase 10.1117/12.458566} {\bibfield  {journal} {\bibinfo  {journal} {Proc. SPIE Int. Soc. Opt. Eng.}\ }\textbf {\bibinfo {volume} {4856}},\ \bibinfo {pages} {90--97} (\bibinfo {year} {2003})}\BibitemShut {NoStop}%
\bibitem [{\citenamefont {Prusti}\ \emph {et~al.}(2016)\citenamefont {Prusti}, \citenamefont {de~Bruijne}, \citenamefont {Brown} \emph {et~al.}}]{Gaia:2016zol}%
  \BibitemOpen
  \bibfield  {author} {\bibinfo {author} {\bibfnamefont {T.}~\bibnamefont {Prusti}}, \bibinfo {author} {\bibfnamefont {J.H.~J.}\ \bibnamefont {de~Bruijne}}, \bibinfo {author} {\bibfnamefont {A.G.~A.}\ \bibnamefont {Brown}},  \emph {et~al.} (\bibinfo {collaboration} {Gaia}),\ }\bibfield  {title} {\enquote {\bibinfo {title} {{The Gaia Mission}},}\ }\href {\doibase 10.1051/0004-6361/201629272} {\  (\bibinfo {year} {2016}),\ 10.1051/0004-6361/201629272},\ \Eprint {http://arxiv.org/abs/1609.04153} {arXiv:1609.04153 [astro-ph.IM]} \BibitemShut {NoStop}%
\bibitem [{\citenamefont {Meiksin}(2006)}]{Meiksin:2005gw}%
  \BibitemOpen
  \bibfield  {author} {\bibinfo {author} {\bibfnamefont {Avery}\ \bibnamefont {Meiksin}},\ }\bibfield  {title} {\enquote {\bibinfo {title} {{Colour corrections for high redshift objects due to intergalactic attenuation}},}\ }\href {\doibase 10.1111/J.1365-2966.2005.09756.X} {\bibfield  {journal} {\bibinfo  {journal} {Mon. Not. Roy. Astron. Soc.}\ }\textbf {\bibinfo {volume} {365}},\ \bibinfo {pages} {807--812} (\bibinfo {year} {2006})},\ \Eprint {http://arxiv.org/abs/astro-ph/0512435} {arXiv:astro-ph/0512435} \BibitemShut {NoStop}%
\bibitem [{\citenamefont {Tortorelli}\ \emph {et~al.}(2024)\citenamefont {Tortorelli}, \citenamefont {McCullough},\ and\ \citenamefont {Gruen}}]{Tortorelli:2024ogu}%
  \BibitemOpen
  \bibfield  {author} {\bibinfo {author} {\bibfnamefont {Luca}\ \bibnamefont {Tortorelli}}, \bibinfo {author} {\bibfnamefont {Jamie}\ \bibnamefont {McCullough}}, \ and\ \bibinfo {author} {\bibfnamefont {Daniel}\ \bibnamefont {Gruen}},\ }\bibfield  {title} {\enquote {\bibinfo {title} {{Impact of stellar population synthesis choices on forward modelling-based redshift distribution estimates}},}\ }\href {\doibase 10.1051/0004-6361/202450694} {\bibfield  {journal} {\bibinfo  {journal} {Astron. Astrophys.}\ }\textbf {\bibinfo {volume} {689}},\ \bibinfo {pages} {A144} (\bibinfo {year} {2024})},\ \Eprint {http://arxiv.org/abs/2405.06009} {arXiv:2405.06009 [astro-ph.CO]} \BibitemShut {NoStop}%
\bibitem [{\citenamefont {Mentasti}\ and\ \citenamefont {Contaldi}(2025)}]{Mentasti:2025ttn}%
  \BibitemOpen
  \bibfield  {author} {\bibinfo {author} {\bibfnamefont {Giorgio}\ \bibnamefont {Mentasti}}\ and\ \bibinfo {author} {\bibfnamefont {Carlo~R.}\ \bibnamefont {Contaldi}},\ }\bibfield  {title} {\enquote {\bibinfo {title} {{A unified spin-harmonic framework for correlating pulsar timing, astrometric deflection, and shimmering gravitational wave observations}},}\ }\href@noop {} {\  (\bibinfo {year} {2025})},\ \Eprint {http://arxiv.org/abs/2507.16906} {arXiv:2507.16906 [astro-ph.CO]} \BibitemShut {NoStop}%
\bibitem [{\citenamefont {Vaglio}\ \emph {et~al.}(2025)\citenamefont {Vaglio}, \citenamefont {Falxa}, \citenamefont {Mentasti}, \citenamefont {Renzini}, \citenamefont {Kuntz}, \citenamefont {Barausse}, \citenamefont {Contaldi},\ and\ \citenamefont {Sesana}}]{Vaglio:2025tex}%
  \BibitemOpen
  \bibfield  {author} {\bibinfo {author} {\bibfnamefont {Massimo}\ \bibnamefont {Vaglio}}, \bibinfo {author} {\bibfnamefont {Mikel}\ \bibnamefont {Falxa}}, \bibinfo {author} {\bibfnamefont {Giorgio}\ \bibnamefont {Mentasti}}, \bibinfo {author} {\bibfnamefont {Arianna~I.}\ \bibnamefont {Renzini}}, \bibinfo {author} {\bibfnamefont {Adrien}\ \bibnamefont {Kuntz}}, \bibinfo {author} {\bibfnamefont {Enrico}\ \bibnamefont {Barausse}}, \bibinfo {author} {\bibfnamefont {Carlo}\ \bibnamefont {Contaldi}}, \ and\ \bibinfo {author} {\bibfnamefont {Alberto}\ \bibnamefont {Sesana}},\ }\bibfield  {title} {\enquote {\bibinfo {title} {{Searching for Gravitational Waves with Gaia and its Cross-Correlation with PTA: Absolute vs Relative Astrometry}},}\ }\href@noop {} {\  (\bibinfo {year} {2025})},\ \Eprint {http://arxiv.org/abs/2507.18593} {arXiv:2507.18593 [gr-qc]} \BibitemShut {NoStop}%
\bibitem [{\citenamefont {Fink}\ \emph {et~al.}(2025)\citenamefont {Fink}, \citenamefont {Contaldi},\ and\ \citenamefont {Mentasti}}]{Fink:2025wlk}%
  \BibitemOpen
  \bibfield  {author} {\bibinfo {author} {\bibfnamefont {Elias}\ \bibnamefont {Fink}}, \bibinfo {author} {\bibfnamefont {Carlo}\ \bibnamefont {Contaldi}}, \ and\ \bibinfo {author} {\bibfnamefont {Giorgio}\ \bibnamefont {Mentasti}},\ }\bibfield  {title} {\enquote {\bibinfo {title} {{Cross-correlating Astrometric and Timing Residuals to Constrain Stochastic Gravitational-Wave Backgrounds}},}\ }\href@noop {} {\  (\bibinfo {year} {2025})},\ \Eprint {http://arxiv.org/abs/2510.25646} {arXiv:2510.25646 [gr-qc]} \BibitemShut {NoStop}%
\bibitem [{\citenamefont {Maggiore}(2007)}]{Maggiore:2007ulw}%
  \BibitemOpen
  \bibfield  {author} {\bibinfo {author} {\bibfnamefont {Michele}\ \bibnamefont {Maggiore}},\ }\href {\doibase 10.1093/acprof:oso/9780198570745.001.0001} {\emph {\bibinfo {title} {{Gravitational Waves. Vol. 1: Theory and Experiments}}}}\ (\bibinfo  {publisher} {Oxford University Press},\ \bibinfo {year} {2007})\BibitemShut {NoStop}%
\bibitem [{\citenamefont {Corless}\ \emph {et~al.}(1996)\citenamefont {Corless}, \citenamefont {Gonnet}, \citenamefont {Hare}, \citenamefont {Jeffrey},\ and\ \citenamefont {Knuth}}]{Corless:1996zz}%
  \BibitemOpen
  \bibfield  {author} {\bibinfo {author} {\bibfnamefont {Robert}\ \bibnamefont {Corless}}, \bibinfo {author} {\bibfnamefont {Gaston}\ \bibnamefont {Gonnet}}, \bibinfo {author} {\bibfnamefont {D.}~\bibnamefont {Hare}}, \bibinfo {author} {\bibfnamefont {David}\ \bibnamefont {Jeffrey}}, \ and\ \bibinfo {author} {\bibfnamefont {D.}~\bibnamefont {Knuth}},\ }\bibfield  {title} {\enquote {\bibinfo {title} {On the lambert w function},}\ }\href {\doibase 10.1007/BF02124750} {\bibfield  {journal} {\bibinfo  {journal} {Advances in Computational Mathematics}\ }\textbf {\bibinfo {volume} {5}},\ \bibinfo {pages} {329--359} (\bibinfo {year} {1996})}\BibitemShut {NoStop}%
\end{thebibliography}%

\end{document}